\documentclass[12pt]{iopart}

\usepackage{amssymb,mathrsfs,bm,mathdots,xspace,bbold,version}   
\usepackage{amsmath}
\usepackage{graphicx}

\usepackage{tikz}

\def\AA{{\cal A}}
\def\CC{{\cal C}}

\def\PP{{\cal P}}
\def\OO{{\cal O}}

\def\hatf{{\hat F}}
\def\hata{{\hat A}}
\def\hatb{{\hat B}}

\def\Go{G^{(0)}}
\def\Gi{G^{(1)}}
\def\Gii{G^{(2)}}
\def\Giii{G^{(3)}}

\def\Sbb{{\mathbb S}}

\def\Beq{\begin{eqnarray}}
\def\Eeq{\end{eqnarray}}

\def\zzz{\hphantom{.}}

\def\moyenne#1{\langle#1\rangle}

\usepackage{soul}

\newcommand{\patho}[1]{%
\begin{tikzpicture}[#1]%
\draw[->] (0,0.0ex) -- (2ex,0.0ex);%
\end{tikzpicture}
}

\newcommand{\pathi}[1]{%
\begin{tikzpicture}[#1]%
\draw (0,0) -- (1ex,0ex);%
\draw (1ex,0ex) -- (1ex,1ex);%
\draw[->] (1ex,1ex) -- (2ex,1ex);
\end{tikzpicture}
}
\newcommand{\pathii}[1]{%
\begin{tikzpicture}[#1]%
\draw (0,0) -- (1ex,0ex);%
\draw (1ex,0ex) -- (1ex,1ex);%
\draw (1ex,1ex) -- (2ex,1ex);%
\draw (2ex,1ex) -- (2ex,0ex);%
\draw[->] (2ex,0ex) -- (3ex,0ex);
\end{tikzpicture}
}

\def\Patho{{\makebox[13pt][l]{\patho{scale=1.25}}}}
\def\Pathos{{\makebox[8pt][l]{\patho{scale=0.85}}}}
\def\Pathi{{\makebox[13pt][l]{\pathi{scale=1.25}}}}
\def\Pathis{{\makebox[8pt][l]{\pathi{scale=0.85}}}}
\def\Pathii{{\makebox[20pt][l]{\pathii{scale=1.25}}}}
\def\Pathiis{{\makebox[15pt][l]{\pathii{scale=0.85}}}}

\begin{document}


\title[Analytical results for stochastic models with arbitrary partitioning of proteins]{Analytical results for a stochastic model of gene expression with arbitrary partitioning of proteins}

\author{Hugo Tschirhart$^{1,2,3}$ \& Thierry Platini$^{4}$}

\address{$^1$ University of Luxembourg, Physics and Materials Science Research Unit, Avenue de la Fa{\"\i}encerie 162a, L-1511 Luxembourg, Luxembourg\\
$^2$ Groupe de Physique Statistique, Institut Jean Lamour (CNRS UMR 7198), Universit\'e de Lorraine Nancy, B.P. 70239, F54506 Vandoeuvre-l\`es-Nancy Cedex, France,\\
$^3$ Doctoral College for the Statistical Physics of Complex Systems, Leipzig-Lorraine-Lviv-Coventry (${\mathbb L}^4$)\\
$^4$ Applied Mathematics Research Center, Coventry University, Coventry, CV1 5FB, England,
}
\ead{thierry.platini@coventry.ac.uk}
\vspace{10pt}
\begin{indented}
\item[]\date{\today}
\end{indented}

\begin{abstract}
In biophysics, the search for analytical solutions of stochastic models of cellular processes is often a challenging task. 
In recent work on models of gene expression, it was shown that a mapping based on {partition}ing of Poisson arrivals (PPA-mapping) can lead to exact solutions for previously unsolved problems. 
While the approach can be used in general when the model involves Poisson processes corresponding to creation or degradation, current applications of the method and new results derived using it have been limited to date. In this paper, we present the exact solution of a variation of the two-stage model of gene expression (with time dependent transition rates) describing the arbitrary partitioning of proteins. The methodology proposed makes full use of the the PPA-mapping by transforming the original problem into a new process describing the evolution of three biological switches. Based on a succession of transformations, the method leads to a hierarchy of reduced models. We give an integral expression of the time dependent generating function as well as explicit results for the mean, variance, and correlation function. Finally, we discuss how results for time dependent parameters can be extended to the three-stage model and used to make inferences about models with parameter fluctuations induced by hidden stochastic variables.
\end{abstract}



%
%
%
%
%

\section{Introduction}
Gene expression is the biological process by which information from a gene is used to synthesize RNA macromolecules and proteins. With a few exceptions, until the 1990s, this process was commonly understood from ``a deterministic viewpoint'' \cite{Ko_1990,Ko_1991}. Since then, the combination of experimental and theoretical approaches has clarified that gene expression is often stochastic in nature (see \cite{Raj_2008,Larson_2009,Huang_2009,Eldar_2010,Lionnet_2012} for review articles). The effect of fluctuations (noise) is usually limited when we are dealing with large numbers of molecules \cite{Van_Kampen}. In cells however, wherein  genes and mRNAs are often present in low numbers, stochasticity has an important role on cellular functions. The importance of fluctuations \cite{Olesen_1999,Becskei_2001,Elowitz_2002,Ozbudak_2002,Blake_2003,Acar_2005,Balazsi_2011} can be illustrated by the observation that, amongst a genetically identical population in a homogenous environment, cell-to-cell variations in gene expression can result in phenotypic heterogeneity.

There exists various mechanisms, some more complex than others, {allow}ing cells to tame and exploit randomness \cite{Nevozhay_2005}. In order to unveil those processes, research efforts are directed on both experimental \cite{Golding_2005,Raj_2006,Taniguchi_2010,Ferguson_2012}
 and theoretical fronts \cite{Peccoud_1995,Hasty_2000,Paulsson_2004,Karmakar_2004,Hornos_2005,Friedman_2006,Okabe_2007,Ramos_2011,Zhang_2012}. Collaborations between biologists, physicists and mathematicians aim to reveal the conditions under which transcriptional noise may, or may not, cascade to affect downstream genetic products.
 
The two stage and three stage models \cite{McAdams_1997,Coulon_2010,Kepler_2001,Paulsson_2005,Paszek_2007} give a minimalist description of the simplest yet non-trivial biological processes leading to gene expression. The two-stage model includes only transcription and translation processes, while the three-stage model also incorporates free and repressed states of the DNA promoter region. Analytical techniques and results \cite{Shahezaei_2008,Biswas_2009,Bokes_2012} for the previously mentioned processes are the cornerstone for further theoretical developments. These models are the elementary bricks {allow}ing for the construction of more complex reaction networks including non-exponential waiting times, transcriptional burst, feedback loops {\it et cetera} \cite{Pedraza_2008,Stinchcombe_2012,Thattai_2001,Xu_2006,Kumar_2014}. Even when it is possible to derive the exact mean and variance of protein and mRNA numbers, obtaining an exact closed-form expression for the generating function is often a challenging problem. The two-stage model is a perfect example. It has been the subject of numerous studies since the paper of Thattai and Oudenaarden \cite{Thattai_2001} in 2001. The model presents linear propensities so that all moments can be derived exactly. Such problems (like the one considered in this paper) are said to be ``exactly solvable''. Nevertheless, the exact generating function for the two stage model \cite{Bokes_2012} was obtained only after ten years of extensive theoretical and experimental studies. 

The search for exact solutions is often challenging because a small variation of a model's definition can {make} analytical results unattainable. Typically methods aiming for a full characterisation of a given process (beyond results for the mean and variance), focus on the master equation approach and its partner equation for the generating function. Once the generating function is obtained, all moments are in principle known: given by successive derivatives. This approach can {provide} insights into different limiting cases and into the behaviour of the distribution in different regions of parameter space. It is important to mention that analytical results (for the probability distribution) have been obtained for a class of models such as monomolecular reactions systems \cite{Jahnke_2007} or deficiency zero networks \cite{Anderson_2010,Anderson_2016}. Unfortunately, once outside these classes there exists no systematic analytical recipe applicable independently of a model's structure. Research efforts are naturally turning towards numerical simulations which though powerful ``bring no intuition into the underlying [...] interactions'' \cite{Thattai_2001}. To {reach} a better understanding one needs to investigate the joint distribution of mRNA and proteins, as well as temporal data, beyond the two-time autocorrelation function. As research {progresses}, emphasis is given to real time measurements with the hope to ``expose the true cell dynamics buried in the average'' \cite{Paulsson_2004}. Nowadays experimental advances {allow} for the count of individual molecules over time \cite{Yu_2006,Suter_2011,Wu_2011,Lionnet_2011} highlighting the need for both time-dependent and steady-state theoretical results .

In recent work, the partitioning of Poisson arrivals \cite{Pendar_2013} was invoked to map Poisson processes to simple biological switches. This method 
is based on the separation of creation events (mRNA creation or protein creation) into independent groups. When applicable, this procedure leads to a mapping between creation/degradation process and a simple two-states biological switch. Applied to the two-stage model, this method led to the time dependent protein distributions \cite{Pendar_2013} using already known results \cite{Peccoud_1995,Raj_2006,Biswas_2009} for mRNA distributions in models with promoter-based regulation. The PPA-mapping needs, however, to be applied with care. It is important to warn the reader that, in a given model, not all creation/degradation process can be mapped onto a biological switch. For the PPA-mapping to apply, one needs to be able to {partition} a given creation event into independent processes. And for a given model, this will depend on upstream regulation of the creation/degradation process under consideration. This restriction is strong and appears as a serious limitation of the mapping applicability. We therefore need a more systematic way to use the PPA-mapping. As it is, the PPA mapping, can only be applied on models presenting a mixture of zero and first order reactions. 
It is unclear as if and how this method can be used or adapted to study models in the presence of feedback. In the simplest model describing bursty mRNA production, a variation of PPA mapping leads to an alternative derivation of the mRNA generating function \cite{Private_Rahul}. But so far, this method has not been the subject of much attention and few models have been solved using this approach.

In direct connection with the applicability question of the PPA mapping is the inverse problem: Assuming the arbitrary partition of a creation event into two `types' (type $1$ and type $2$), the latter process being itself regulated by an upstream {mechanism}, what correlation (between $1$ and $2$) should we expect? Is the correlation bounded? Does it vanish under particular conditions? Also, it is of experimental interest, to search for ways to infer the protein levels of a given type using measurement data on the other.

The model and methodology proposed in this paper were designed to (1) study correlations induced by the arbitrary {partition} of proteins arrival and (2) obtain the generating function making full use of the idea implicit in the PPA-mapping. The process we consider appears to be a simplified version of mechanisms involved in alternating splicing processes allowing a single gene to code for multiple proteins. With alternating splicing, a particular pre-mRNA can lead to different messenger RNAs, each being responsible for the production of isoform proteins (differing in their amino acid sequence). Results recently published in \cite{Wang_2014} focused on both bursty and constitutive pre-mRNA creation. One should mention that alternating splicing is far from being rare. Many genes have multiple splicing patterns \cite{Modrek_2002,Black_2000,Graveley_2001} and numerous examples confirm that alternating splicing contributes to the development of cancer 
(see \cite{Grabowski_2001} and \cite{Black_2003} for review articles). Our goal is to obtain the time dependent solution of the proposed model in term of the generating function. Our method is based on the construction of different mappings. Each transformation aims to reduce the study of a given model to the analysis of a simpler one. After a succession of transformations the problem is condensed to the study of two-state biological switches. The nesting between models and reduced models is reflected in a set of relations between generating functions. This hierarchy allows us to derive relations between mean numbers and higher order moments. We show that the PPA mapping {allow}s us to consider arbitrary time dependent transition rates. Other studies such as \cite{Mugler_2010, Jedrak_2016} and \cite{Dattani_2017} have considered explicit time dependent parameters to investigate the effect of upstream hidden dynamics on downstream populations. Here, we obtain the time dependent generating function without solving any complicated differential equation but rather a simple first order equation (with time dependent coefficients). Accessing analytical results for time dependent coefficient provides a way to tackle models with noisy transition rates induced by hidden stochastic variables. In this paper, we show how results for the mean and correlations for time dependent model can be used to access the solution for the three-stage model.

The paper is organised as follows. In section \eqref{The model}, we start with the presentation of the model under consideration. We give the master equation governing the evolution of the probability distribution and present the solution of the first order moments. In section \eqref{The generating function}, we define the generating function, and outline the three different steps defining our method. 
Step $1$ and $3$ both describe the transformation under the PPA-mapping at different stages of the derivation. The intermediate step $2$ defines
the decomposition of a given process over all possible histories. Each step is detailed in sections \eqref{From 0 to 1}, \eqref{From 1 to 2} and \eqref{From 2 to 3}. The succession of transformations takes us relatively far away from the solution of the original problem. To proceed further, we give in section \eqref{Towards the generating function}, the probability associated to each relevant histories. Finally, the time dependent generating function is presented in section \eqref{Final expression}, where our result is generalised to arbitrary {partition} numbers. We show how for some particular cases our results match the known solution for the two-stage model (see reference \cite{Bokes_2012} and \cite{Pendar_2013}). Finally we discuss how to use results for time dependent parameters to extend this work to the three-stage model and others processes including additional random variables.
\section{Theory}
\subsection{The model}\label{The model}
\begin{figure}
  \centering
  \includegraphics[width=0.5\linewidth]{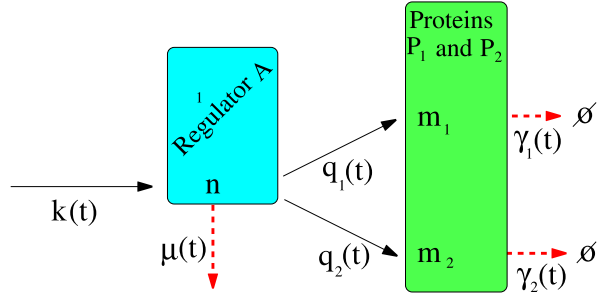}\\
  \caption{\label{FIG1} The original model (model-0): Two types of proteins ($\PP_1$ and $\PP_2$) are regulated by an upstream molecule $\AA$. Transition rates for protein production and degradation are respectively written $q_j(t)$ and $\gamma_j(t)$ (with $j=1,2$), while $k(t)$ and $\mu(t)$ denote production and degradation rates for regulator $\AA$.} 
\end{figure}

The model under consideration describes the stochastic evolution of protein numbers in a variation of the two-stage model, for which proteins are arbitrarily separated into two groups ($\PP_1$ and $\PP_2$). Note that our results will be easily generalized to an arbitrary number of protein types. We denote by $\AA$ the upstream molecule regulating proteins production (see Figure \eqref{FIG1}). Time dependent transition rates for proteins production and degradation are respectively written $q_j(t)$ and $\gamma_j(t)$ (with $j=1,2$). The level of regulator $\AA$ is itself governed by the transition rates $k(t)$ and $\mu(t)$, respectively associated to creation and degradation (see table \eqref{table_mod1}). Because the method presented in this paper invokes mappings to other processes, it is convenient to refer to the original model as model-0. 

The state of the system is, at any time, characterised by the numbers $n$, $m_1$ and $m_2$, of molecules $\AA$ and proteins $\PP_1$ and $\PP_2$ respectively. We write $P_{n,m_1,m_2}(t)$ the probability distribution of state $(n,m_1,m_2)$. We should keep in mind that the latter quantity is conditional on the initial state. In particular, we will consider $P_{0,0,0}(t=0)=1$. Further down in this paper we will explain how the latter initial condition is imposed by the initial state of the reduced model resulting from successive mappings. In order to consider other initial states, an extension of the proposed method is needed. This procedure would add an extra layer of complexity to the work presented here and is not discussed further in this paper. The probability distribution is governed by the master equation
\begin{eqnarray}
\label{eq_Master_eq}
\frac{d}{dt}P_{n,m_1,m_2} &=&  k(t)[P_{n-1,m_1,m_2}- P_{n,m_1,m_2}]\\
&+& \mu(t)[(n+1) P_{n+1,m_1,m_2}- n P_{n,m_1,m_2}]\nonumber\\
&+& q_1(t)n[ P_{n,m_1-1,m_2} - P_{n,m_1,m_2}] \nonumber \\
&+& q_2(t)n[ P_{n,m_1,m_2-1} - P_{n,m_1,m_2}]\nonumber \\
&+& \gamma_1(t) [(m_1+1) P_{n,m_1+1,m_2} - m_1 P_{n,m_1,m_2}]\nonumber \\
&+& \gamma_2(t) [(m_2+1) P_{n,m_1,m_2+1} - m_2 P_{n,m_1,m_2}]
.\nonumber
\end{eqnarray}
At this point, it is premature to solve the full master equation, and we start by deriving equations for mean population numbers. Let us denote $\moyenne{\OO}$ the average of the observable $\OO$, given by $\sum_{n,m_1,m_2}\OO(n,m_1,m_2)P_{n,m_1,m_2}$. For mean numbers we derive the equations:
\Beq
\label{eq_2}
\frac{d\moyenne{n}}{dt}&=&k(t)-\mu(t)\moyenne{n},\\
\label{eq_3}
\frac{d\moyenne{m_j}}{dt}&=&q_j(t)\moyenne{n}-\gamma_j(t)\moyenne{m_j}, \ \ j=1,2.
\Eeq
If we restrict ourself to the well known case of constant reaction rates, with $\moyenne{n}(t=0)=\moyenne{m_j}(t=0)=0$, the solutions of these equations are:
\begin{eqnarray}
\label{esay_one}
\frac{\moyenne{n}(t)}{\moyenne{n}^* }&=&\left(1-e^{-\mu t}\right),
\end{eqnarray}
and
\begin{equation}
\frac{\moyenne{m_j}(t)}{\moyenne{m_j}^*}=
\label{eq5}
\left\{
\begin{array}{cc}
1-\frac{1}{\gamma_j-\mu}(\gamma_je^{-\mu t}-\mu e^{-\gamma_j t}), & \gamma_j/\mu\ne1\\
1-(1+\mu t)e^{-\mu t},\ \gamma_j/\mu=1,
\end{array}
\right.
\end{equation}
with the stationary values
$\moyenne{n}^*={k}/{\mu}$ and $\moyenne{m_j}^*={kq_j}/{\mu\gamma_j}$. At this stage, it is not hard to extend these results to time-dependent coefficients. With a bit of work, one can show that the solution of equation $\eqref{eq_2}$ is
\Beq
\label{solution_n_time_dep}
\moyenne{n}(t)=\int_0^t \mathrm{d}s \ k(s)e^{W_\mu(s)-W_\mu(t)},
\Eeq
where we define $W_\mu(t)=\int_0^t\mathrm{d}\lambda \ \mu(\lambda)$. For arbitrary functions $k(t)$ and $\mu(t)$ it is however impossible to comment on the existence of a stationary state unless we assume the existence of the limits $k(t\rightarrow\infty)=k^*$ and $\mu(t\rightarrow\infty)=\mu^*$. It is here important to note that the solution of equation $\eqref{eq_3}$ can be expressed as a double integral. To proceed we use equation $\eqref{solution_n_time_dep}$ with the substitution $k(t)\rightarrow q_j(t)\moyenne{n}(t)$ and $\mu(t)\rightarrow\gamma_j(t)$. A few lines of calculation leads to
\Beq\label{eqKK}
\moyenne{m_j}(t)
&=&\int_0^t \mathrm{d}s \int_{s}^t \mathrm{d}s' \ {\cal K}_j(s,s',t),
\Eeq
with kernel 
\Beq
\label{eq_kernel}
{\cal K}_j(s,s',t)=k(s)q_j(s')e^{W_\mu(s)-W_\mu(s')+W_{\gamma_j}(s')-W_{\gamma_j}(t)},
\Eeq
where $W_{\gamma_j}(t)=\int_0^t\mathrm{d}\lambda \ \gamma_j(\lambda)$. Interestingly, it is this quantity ${\cal K}_j(s,s',t)$ which will reappear explicitely in the final expression for the generating function. Along the same lines, it is possible to push further, writing equations for second order moments such as $\moyenne{n^2}$  \begin{eqnarray} \label{eq_7} \frac{d\moyenne{n^2}}{dt}&=&k(t)+(2k(t)+\mu(t))\moyenne{n}-2\mu(t)\moyenne{n^2}. \end{eqnarray} Once again, the solution for constant coefficients is easy to derive  
and can be expressed as a function of $\moyenne{n}(s)$:
\Beq
\label{solution_n2_time_dep}
\moyenne{n^2}(t)
=\int_0^t \mathrm{d}s \ [k(s)+(2k(s)+\mu(s))\moyenne{n}(s)]e^{2W_\mu(s)-2W_\mu(t)}.
\Eeq
To evaluate correlations of the form $\moyenne{m_1m_2}(t)$ and $\moyenne{nm_1}(t)$ we write
\begin{eqnarray}
\frac{d\moyenne{nm_j}}{dt}&=&k(t)\moyenne{m_j}(t)+q_j(t)\moyenne{n^2}-(\mu(t)+q_j(t))\moyenne{nm_j},\label{eqaa}
\\
\frac{d\moyenne{m_1m_2}}{dt}&=&q_1(t)\moyenne{nm_2}+q_2(t)\moyenne{nm_1}-(\gamma_1(t)+\gamma_2(t))\moyenne{m_1m_2}.\label{eqbb}
\end{eqnarray}
Together with $d\moyenne{m_j}/dt$ ($j=1,2$), $d\moyenne{n}/dt$ and $d\moyenne{n^2}/dt$, Eq. \eqref{eqaa} and \eqref{eqbb} define a system of seven equations. Importantly, equations governing the evolution of correlators do not involve higher order moments. As a consequence, correlations at any order can be obtained by solving a finite set of equations \cite{Lesta_2008, Gadgil_2005, Singh_2007}. Even if one considers constant reaction rates, the generalisation of the solution to three or more protein types is not trivial. It requires the solution of a new and bigger set of equations. One possible avenue is to pursue with approximation of the ``mean field'' type, which consists in assuming $\moyenne{m_1m_2}\simeq\moyenne{m_1}\moyenne{m_2}$. A priori, the later approximation holds for weakly correlated systems only. Hence we need to quantify correlation numbers in order to select the appropriate analytical methods.
\begin{table}[!h]
\centering
   \begin{tabular}{c c c}
   \hline
    Event	& Update & Transition rates  \\ \hline	\hline & \\
    $\AA$-production  & $n\rightarrow n+1$ & $k(t)$ \\ \\ 
    $\AA$-degradation & $n\rightarrow n-1$ & $n\mu(t) $ \\ \\ \hline \hline & \\
    $\PP_j$-production & $m_j\rightarrow m_j+1$ & $n q_j(t)$ \\ \\
    $\PP_j$-degradation & $m_j\rightarrow m_j-1$ & $m_j\gamma_j(t)$ \\ \\
\hline
	\end{tabular}
  \caption{Transitions and associated rates for the original model (model-0).}
  \label{table_mod1}
\end{table}
\subsection{The generating function}\label{The generating function}
\begin{figure}[h!]
  \centering
  \includegraphics[width=0.5\linewidth]{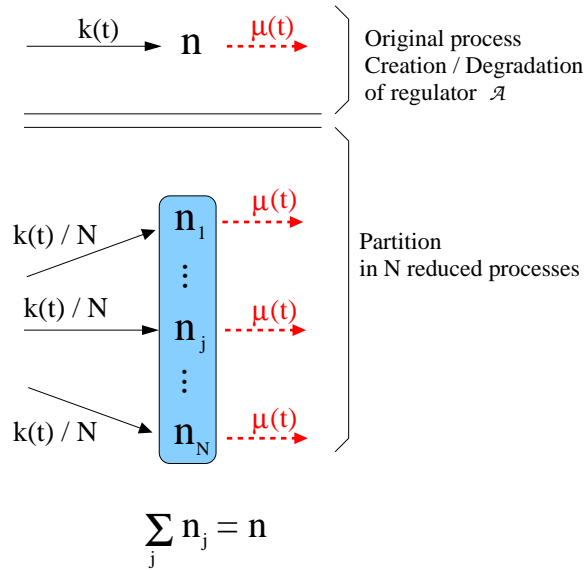}\\
  \caption{    \label{FIG00} We {partition} each creation event into $N$ 'types'. The creation rate associated to a particular type is given by $k(t)/N$. The sum of molecules numbers $n_j$ over each type is equal to the total number of molecules $n$ in the original model.} 
\end{figure}

Let us start by defining the generating function of the original model
\Beq
\Go(x,z_1,z_2,t)=\sum_{n,m_1,m_2}x^nz_1^{m_1}z_2^{m_2}P_{n,m_1,m_2}(t),
\Eeq
which obeys the differential equation
\Beq
\label{eq_dG/dt}
\frac{d\Go}{dt}&=&(x-1)(k(t)-\mu(t)\partial_{x})\Go\\
&+&(z_1-1)(q_1(t)x\partial_{x} - \gamma_1(t)\partial_{z_1})\Go+
(z_2-1)(q_2(t)x\partial_{x} - \gamma_2(t)\partial_{z_2})\Go\nonumber.
\Eeq
Focusing our attention on the numbers of proteins only, we define the marginal probability
\Beq
P_{m_1,m_2}(t)=\sum_{n=0}^\infty P_{n,m_1,m_2}(t),
\Eeq
for which the generating function is $G^{(0)}(z_1,z_2,t)=G^{(0)}(1,z_1,z_2,t)$. In order to attain an analytical expression we will successively reduce the original model into simpler ones. To avoid confusion we choose to denote as model-1, model-2 and model-3, the processes which will be emerging from these successive mappings. We write $\Gi$, $\Gii$ and $\Giii$ the generating functions for each model respectively. The following gives a short description of the steps taken in this paper, while each of them is further developed in sections \eqref{From 0 to 1}, \eqref{From 1 to 2} and \eqref{From 2 to 3}.
\begin{enumerate}
\item[Step 1:] The PPA-mapping \cite{Pendar_2013} is based on the {partition}ing of Poisson processes (see Figure \eqref{FIG00}). It {allow}s for simplification of the original problem to $N$ independent processes all identical to model-1 (see Figure \eqref{FIG2} and section \eqref{From 0 to 1}). In the reduced model, the production of protein is regulated by a biological switch taking values $\theta=0$ (OFF) and $\theta=1$ (ON). Since $N$ appears as a parameter of the reduced model, we write $\Gi_N(z_1,z_2,t)$ the generating function of model-1. The latter is related to the original generating function via:
\Beq
G^{(0)}(z_1,z_2,t)=\lim_{N\rightarrow\infty}{\Big[}\Gi_N(z_1,z_2,t){\Big]}^N.
\label{eq_G_g}
\Eeq
\item[Step 2:] Denoting by $\Theta$ a particular history (or path) generated by the time evolution of the variable $\theta$, we define $\Psi_N(\Theta)$ to be the probability of a given path. Model-2 is defined for one particular history as if frozen (Figure \eqref{FIG3}). We write $\Gii_\Theta$ as the associated generating function and express $\Gi_N$ as an average over all possible histories (see section \eqref{From 1 to 2})
\Beq
\Gi_N(z_1,z_2,t)=\sum_\Theta \Psi_N(\Theta)\Gii_{\Theta}(z_1,z_2,t).
\label{eq_g_H}
\Eeq
Once the differential equation for $\Gii_{\Theta}$ has been derived, we will be able to show that protein numbers are uncorrelated in model-2. It follows that $\Gii_\Theta$ can be expressed as the product of two functions, each associated to a given protein type:
\Beq
\Gii_{\Theta}(z_1,z_2,t)=\prod_{j=1,2}\Gii_{j|\Theta}(z_j,t).
\Eeq
\item[Step 3:] To access the solution of model-2, we exploit the PPA-mapping one more time. Splitting the creation process into $M$ independent processes, it ultimately reduces to the study of biological switches (see model-3 in Figure \eqref{FIG4}). Writing $\Giii_{M;j|\Theta}(z_j,t)$ as the generating function of the switch $j$ ($j=1,2$), we show the relation (see section \eqref{From 2 to 3})
\end{enumerate}
\Beq
\label{eqxxx}
\Gii_{j|\Theta}=\lim_{M\rightarrow\infty}\left[\Giii_{M;j|\Theta}\right]^M.
\Eeq
Finally, nesting all steps together, the original generating function is given by
\Beq
G^{(0)}(z_1,z_2,t)=
\lim_{N\rightarrow\infty}{\Bigg[}
		\underbrace{
			\sum_\Theta \Psi_{N}(\Theta)
			\underbrace{
				\prod_{j=1,2}
				\underbrace{\lim_{M\rightarrow\infty}{\Big[}
\Giii_{M;j|\Theta}(z_j,t)
{\Big]}^M}_{\Gii_{j|\Theta}(z_j,t)}
			}_{\Gii_\Theta(z_1,z_2,t)}
		}_{\Gi_{N}(z_1,z_2,t)}
		{\Bigg]}^N.
\Eeq 
\subsection{Consequences: hierarchy in mean and correlation numbers} \label{Hierarchy in numbers}
Before entering the heart of the subject with the application of the PPA-mapping, one can investigate consequences of these successive transformations. The nesting of generating functions {allow}s us to derive direct relations between mean numbers in the different models. We write $\moyenne{m_j}_N^{(1)}$, $\moyenne{m_j}_{\Theta}^{(2)}$ and $\moyenne{m_j}_{M|\Theta}^{(3)}$ the mean numbers of $j$-proteins in model-1, 2 and 3 respectively. For simplicity, we choose to omit the superscript $0$ so that $\moyenne{m_j}$ denotes the average number of proteins in the original model. To ease the notations further we choose not to make the time dependance explicit, since the relations derived bellow are true for all time $t$. Equations \eqref{eq_G_g}, \eqref{eq_g_H} and \eqref{eqxxx} bring us to
\Beq
\label{eqh1}
\moyenne{m_j}&=&\lim_{N\rightarrow\infty}N\moyenne{m_j}_N^{(1)},\\
\label{eqh2}
\moyenne{m_j}_N^{(1)}&=&\sum_{\Theta}\Psi_N(\Theta)\moyenne{m_j}^{(2)}_{\Theta},\\
\label{eqh3}
\moyenne{m_j}^{(2)}_{\Theta}&=&
\lim_{M\rightarrow\infty}M\moyenne{m_j}_{M|\Theta}^{(3)}.
\Eeq
The calculation of $\moyenne{m_j}^{(3)}$ is a pretty simple affair. Each protein being reduced to a biological switch, $m_j^{(3)}$ is restricted to the value $0$ and $1$. We give here, the expression of $\moyenne{m_j}^{(3)}$, for which the derivation is presented in section \eqref{From 2 to 3}:
\Beq
\moyenne{m_j}_{M|\Theta}^{(3)}=\frac{1}{M}\int_0^t \mathrm{d}\lambda \ \Theta(\lambda)q_j(\lambda)e^{W_{\gamma_j}(\lambda)-W_{\gamma_j}(t)}.
\Eeq
To continue further, eq. \eqref{eqh2} requires knowledge of the probability $\Psi_N(\Theta)$ for a given path. This is not particularly difficult as one only needs to consider paths with probability up to the order $1/N$ (see section \eqref{Towards the generating function}). Without further knowledge of the generating function, once $\Psi_N(\Theta)$ and $\moyenne{m_j}^{(3)}$ given, the reader can derive the time evolution for mean number of proteins using equations \eqref{eqh1}, \eqref{eqh2} and \eqref{eqh3}. Practically, those steps give a convoluted way to reach the result already presented in \eqref{eq5}. It however reflects on the strategy adopted here to access the generating function. \\
Considering the correlation function, with the help of Eq. \eqref{eq_G_g}, we can show 
\Beq
C_{1,2}=\moyenne{m_1m_2}-\moyenne{m_1}\moyenne{m_2}=\lim_{N\rightarrow\infty}N\moyenne{m_1m_2}^{(1)}_N.
\Eeq
The protein number being uncorrelated in model-2 we have $\moyenne{m_1m_2}^{(2)}_{\Theta}=\moyenne{m_1}_{\Theta}^{(2)}\moyenne{m_2}_{\Theta}^{(2)}$, which leads us to
\Beq
\moyenne{m_1m_2}_N^{(1)}=\sum_{\Theta}\Psi_N(\Theta)\moyenne{m_1}_{\Theta}^{(2)}\moyenne{m_2}_{\Theta}^{(2)}.
\Eeq
From the latter two equations, we conclude that $C_{1,2}>0$ unless at least one of $\moyenne{m_j}=0$ ($j=1,2$). Hence, there is no non-trivial point in parameter space such that the correlation between protein number vanishes. As a consequence, there is no region of the parameter space in which the mean field approach is valid. In \cite{Wang_2014}, the authors focus on alternative splicing mechanism, investigating the stationary state of a slightly different model from the one presented here. This study considers the transition from a pre-mRNA to two different mature mRNAs. For constitutive expression (no bursty pre-mRNA creation), they show that (in the stationary state) the mRNA numbers (of type $1$ and $2$) are independent. They however observe, for bursty pre-mRNA production, the emergence of correlations between the two mature mRNA types. 

\subsection{First transformation: from model-0 to model-1}\label{From 0 to 1}
\begin{figure}
  \centering
  \includegraphics[width=0.5\linewidth]{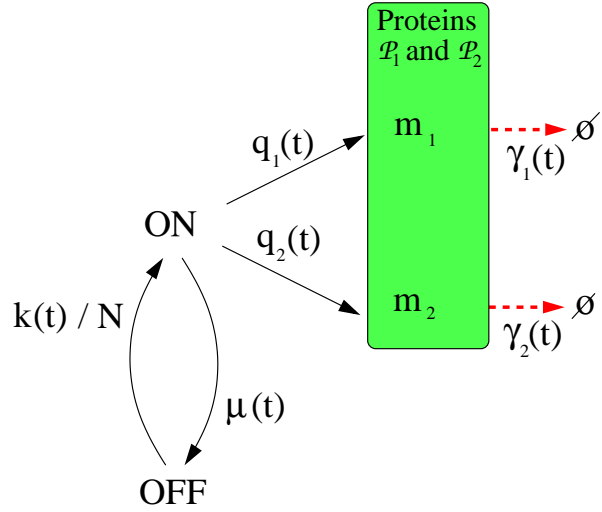}\\
  \caption{ \label{FIG2}
  Model-1: The creation and degradation of two types of proteins ($\PP_1$ and $\PP_2$) is regulated by an upstream switch. Transition rates for proteins production and degradation are respectively written $q_j(t)$ and $\gamma_j(t)$ (with $j=1,2$), while $k(t)/N$ and $\mu(t)$ denote the probabilities of transition from $OFF\rightarrow ON$ and $ON\rightarrow OFF$.  } 
\end{figure}

\begin{figure}
  \centering
  \includegraphics[width=0.5\linewidth]{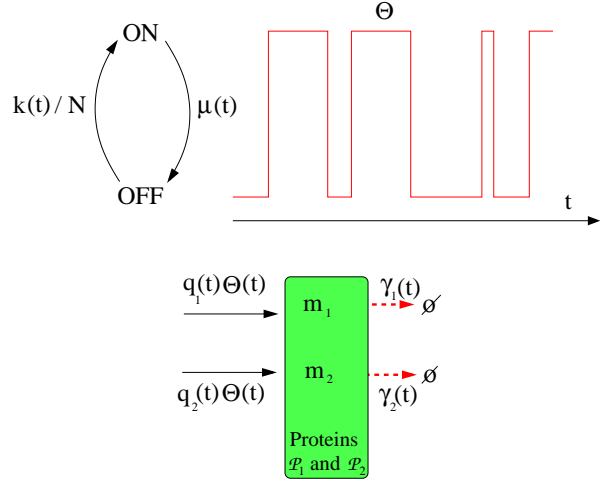}\\
  \caption{ \label{FIG3}Model-2: The creation and degradation of two types of proteins ($\PP_1$ and $\PP_2$) for a fixed history $\Theta$. Transition rates for proteins production and degradation are respectively written $q_j(t)\Theta(t)$ and $\gamma_j(t)$ (with $j=1,2$).  } 
\end{figure}

The PPA-mapping is based on the {partition}ing property of Poisson processes. Without entering into technical details, the mapping can be understood as follow: 
\begin{enumerate}
\item Consider the creation/degradation process of regulator $\AA$ (with rates $k(t)$ and $\mu(t)$). 
\item Partition every creation events into $N$ 'types' (Figure \eqref{FIG00}). The {partition} is homogeneous so that each $\AA$ molecule is equally likely to be assigned to a given type. It follows that the creation rate associated to a particular type is given by $k(t)/N$. 
\item Take the limit $N\gg1$. As a consequence, the probability of observing more than one $\AA$ molecule of a particular type can be neglected. It follows that the random variable describing the number of molecules $\AA$ (of a given type) is restricted to the value $0$ or $1$.
\end{enumerate}
Model-1 as defined under this procedure is illustrated on Figure \eqref{FIG2}. Note that $N$ appears as a parameter in the reduced model. Along the lines presented in \cite{Pendar_2013} we write $G^{(0)}=[\Gi_N]^N$. Equation \eqref{eq_dG/dt} shows that $\Gi_N$ obeys the same differential equation under the transformation $k(t)\rightarrow k(t)/N$. As a consequence the probability of observing (in the reduced model) more than one $\AA$ molecule is of order $1/N^2$ and can be neglected as $N\rightarrow\infty$. While the previous logical argument shows how model-1 is emerging from model-0, an alternative derivation, based on the probability distribution instead of the generating function, {allow}s for the reversed construction: building model-0 starting with $N$ independent model-1. This derivation, not presented in the literature so far, is presented in an appendix.
\subsection{Second transformation: from model-1 to model-2}\label{From 1 to 2}
Let us remind the reader that $\theta$ is the new stochastic variable (taking value in $\{0,1\}$) emerging in model-1. The decomposition over all possible histories, generated by the variable $\theta$, emerges from the use of conditional probabilities. To be more explicit we write $\Theta$ as a particular history associated to the variable $\theta$. For a given path, we write $\Theta(t)$ as the value taken by the random variable $\theta$ at time $t$. We continue further by writing $\varphi_{N;(\Theta,a,b)}(t)$ as the probability associated to a particular history $\Theta$ and protein numbers $a$ and $b$. The generating function $\Gi_N$ can be rewritten as
\Beq
\Gi_N(z_1,z_2,t)=\sum_{\Theta,a,b}z_1^az_2^b\varphi_{N;(\Theta,a,b)}(t).
\Eeq
Defining $\psi_{a,b|\Theta}(t)$ as the conditional probability on $\Theta$ while $\Psi_N(\Theta)$ is the probability of a given history, the equality $\varphi_{N;(\Theta,a,b)}(t)=\Psi_N(\Theta)\psi_{a,b|\Theta}(t)$ leads to
\Beq
\Gi_N(z_1,z_2,t)=\sum_{\Theta}\Psi_N(\Theta)\Gii_{\Theta}(z_1,z_2,t),
\Eeq
with
\Beq
\Gii_{\Theta}(z_1,z_2,t)=\sum_{a,b}z_1^az_2^b\psi_{a,b|\Theta}(t).
\Eeq
For a known history $\Theta(t)$, we have 
\Beq
\frac{d\Gii_\Theta}{dt}&=&(z_1-1)(q_1(t)\Theta(t)-\gamma_1(t)\partial_{z_1})\Gii_\Theta\\
&+&(z_2-1)(q_2(t)\Theta(t)-\gamma_2(t)\partial_{z_2})\Gii_\Theta.
\nonumber
\Eeq
Note that in the last equation $\Gii_\Theta$ is clearly independent of $N$ and so is the conditional probability $\psi_{a,b|\Theta}$. The dependence in $N$ is now carried by the probability $\Psi_N$. At this point we see that $\Gii_\Theta$ can be written as
\Beq
\Gii_\Theta(z_1,z_2,t)=\prod_{j=1,2}\Gii_{j|\Theta}(z_j,t),
\Eeq
where each generating function is governed by
\Beq
\label{eqG2j}
\frac{d \Gii_{j|\Theta}}{dt}=(z_j-1)(q_j(t)\Theta(t)-\gamma_j(t)\partial_{z_j})\Gii_{j|\Theta}.
\Eeq
Thereupon the two protein numbers are uncorrelated in model-2. 
\subsection{Third transformation: from model-2 to model-3}\label{From 2 to 3}
\begin{figure}
  \centering
  \includegraphics[width=0.5\linewidth]{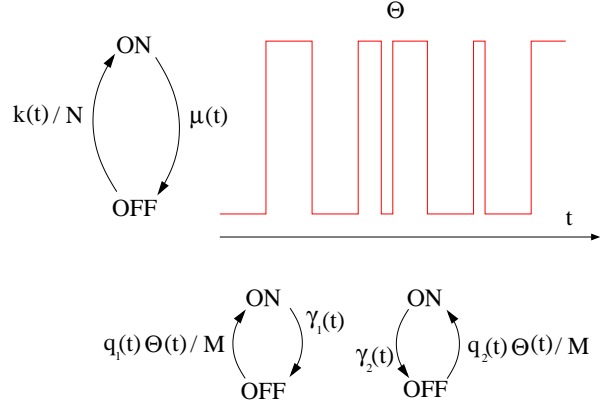}\\
  \caption{\label{FIG4} Model-3 describes, for a fixed history $\Theta$, two biological switches for which transition rates are respectively written $q_j(t)\Theta(t)/M$ and $\gamma_j(t)$ (with $j=1,2$).} 
\end{figure}

To {reach} the expression of $\Gii_\Theta$, one applies the PPA-mapping one more time. This will reduce the original model to the study of biological switches (Figure \eqref{FIG4}). For each protein type $\PP_j$ ($j=1,2$), we once again, choose to {partition} every creation event into $M$ groups. The {partition} being homogeneous, each protein is equally likely to be assigned to a given group. The creation rate for a particular group is given by $q_j(t)/M$. Taking the limit $M\gg1$ {allow}s us to neglect the creation of more than one protein in each group. To be explicit one writes $\Gii_{j|\Theta}=[\Giii_{M;j|\Theta}]^M$ in equation \eqref{eqG2j}. This procedure leads to the same differential equation with the transformation $q_j(t)\Theta(t)\rightarrow q_j(t)\Theta(t)/M$. Hence, in the limit $M\rightarrow\infty$, the number of proteins of type $j$ are restricted to $0$ and $1$. It follows that
\Beq
\label{eq23}
\Gii_\Theta=\prod_{j=1,2}\lim_{M\rightarrow\infty}{\Big[}
\Giii_{M;j|\Theta}
{\Big]}^M.
\Eeq
The function $\Giii_{M;j|\Theta}$ describes the dynamics of a two-state model and can be written has
\Beq
\label{eqh^j}
\Giii_{M;j|\Theta}(z,t)=1+(z-1)f_{M;j|\Theta},
\Eeq 
where $f_{M;j|\Theta}$ is the probability to find the switch $j$ in the ON-state, knowing the history $\Theta$. The latter is the solution of the following equation
\Beq
\frac{d f_{M;j|\Theta}}{dt}=q_j(t)\Theta(t)/M-\left[\gamma_j(t)+q_j(t)\Theta(t)/M\right]f_{M;j|\Theta}.
\Eeq
We now have reached the point where one needs to define the initial state. We choose to consider $f_{M;j|\Theta}(t=0)=0$. Let us remind the reader that, in model-3, the total number of switches $j$ in the ON-state equals the number of proteins $\PP_j$ in model-2. The hierarchy builds up to the number of proteins in model-0. As we look at equation \eqref{eqh1}, \eqref{eqh2} and \eqref{eqh3}, we see that choosing (at time $t=0$) all switches ($j=1,2$) in the OFF-state, imposes the following initial state on to the original model
\Beq
\label{initial_state_1}
m_1(t=0)=m_2(t=0)=0.
\Eeq
A simple calculation gives
\Beq
f_{M;j|\Theta}(t) = 
\int_0^t\mathrm{d}\lambda \ \frac{q_j(\lambda)}{M}\Theta(\lambda) 
\exp\left[-\int_\lambda^t \mathrm{d} s\ \left\{\gamma_j(s)+\frac{q_j(s)}{M}\Theta(s)\right\}\right],
\Eeq
which, to the first order in $1/M$, simplifies to
\Beq
f_{M;j|\Theta}(t) &=&{\Lambda_{j|\Theta}(t)}/{M},
\label{eqf}
\Eeq
with 
\Beq
\Lambda_{j|\Theta}(t)=\int_0^t\mathrm{d}\lambda \ \Theta(\lambda)q_j(\lambda)e^{W_{\gamma_j}(\lambda)-W_{\gamma_j}(t)}.
\label{eqLAMBDA}
\Eeq
Nesting equation \eqref{eqf} into \eqref{eqh^j} leads to:
\Beq
\Giii_{M;j|\Theta}(z,t)=1+(z-1){\Lambda_{j|\Theta}(t)}/{M}.
\Eeq
With Eq. \eqref{eq23} the latter result {allow}s us to write
\Beq
\label{eqH}
\Gii_\Theta(z_1,z_2,t) =\exp
\left[
\sum_{j=1,2}(z_j-1)\Lambda_{j|\Theta}(t)
\right].
\Eeq
\subsection{Summing over all histories}\label{Towards the generating function}
In order to derive the generating function $\Gi_N$ from $\Gii_\Theta$, using equation \eqref{eq_g_H}, we focus on the expression of the probability $\Psi_{N}(\Theta)$ for all relevant histories $\Theta$. As mentioned earlier, we simply need to evaluate $\Psi_N(\Theta)$ up to the order $1/N$. We choose to consider the initial state $\Theta(t=0)=0$. Amongst $N$ identical models, the number of switches in the ON-states defines the number $n$ of molecules $\AA$. It follows that the initial state must satisfy:
\Beq
\label{initial_state_2}
n(t=0)=0.
\Eeq
Together, equations \eqref{initial_state_1} and \eqref{initial_state_2} fully specify the initial state, so that $P_{0,0,0}(t=0)=1$. We remind the reader that the probability of the transition form $ON\rightarrow OFF$ (between time $t$ and $t+\delta t$) is given by $\mu(t) \delta t$. In addition, the probability of observing the transition $OFF\rightarrow ON$ is given by $k(t)\delta t/N$. At the first order in $1/N$, three different types of histories are relevant. They are symbolically represented by $\Patho$, $\Pathi$ and $\Pathii$ and detailed in  table \eqref{table_Path}.
The probability associated to each path is
\Beq
& &\Psi_{N}
	\left(
		\Patho
	\right)
		\simeq 1 - \frac{1}{N}\int_0^t\mathrm{d}s\ k(s),	\\
& &	\Psi_{N}
		\left(
		\Pathi
		\right)
		\simeq  \frac{k(s)}{N}e^{W_\mu(s)-W_\mu(t)},  \\
& &	\Psi_{N}
		\left(
		\Pathii
		\right)
		\simeq\frac{k(s)}{N}\mu(s') e^{W_\mu(s)-W_\mu(s')}.
\Eeq
It is particularly useful to rewrite the last equation as
\Beq
\Psi_{N}
		\left(
		\Pathii
		\right)
		\simeq -\frac{k(s)}{N}e^{W_\mu(s)}\left(\partial_{s'}e^{-W_\mu(s')}\right),
\Eeq
which can be used to verify the conservation of probability:
\begin{eqnarray}
		\Psi_{N}
		\left(
		\Patho
		\right)
		+
		\int_{0}^t \mathrm{d}s \ \zzz
		\Psi_{N}
		\left(
		\Pathi
		\right)	
	+		\int_{0}^t \mathrm{d}s
		\int_{s}^t \mathrm{d}s' \
		\Psi_{N}
		\left(
		\Pathii
		\right)=1.
\end{eqnarray}
The latter relation confirms that all relevant paths have been taken into consideration. The expression of $\Gi_N$ is symbolically given by
\begin{eqnarray}
	 \Gi_{N}&=&
		\Psi_{N}
		\left(
		\Patho
		\right)
		\Gii_{\Pathos}
		+
		\left.
		\int_{0}^t \mathrm{d}s \
		\Psi_{N}
		\left(
		\Pathi
		\right)
		\Gii_{\Pathis}
		\right. 	 
		+
		\int_{0}^t \mathrm{d}s
		\int_{s}^t \mathrm{d}s' \
		\Psi_{N}
		\left(
		\Pathii
		\right)
		\Gii_{\Pathiis}.\nonumber\\
\end{eqnarray}
The explicit calculation (using $\Gii_{\Pathos}=1$) points us to
\Beq
\label{eqGiF}
 \Gi_{N}(z_1,z_2,t) \simeq 1 &+& \frac{A(z_1,z_2,t)}{N},
\Eeq
with 
\Beq
A(z_1,z_2,t)=
\int_0^t \mathrm{d}s \ k(s) e^{W_\mu(s)} \int_s^t \mathrm{d}s' \ e^{-W_\mu(s')} \partial_{s'}\Gii_{\Pathiis}.
\Eeq
\begin{table}[!h]
\centering
   \begin{tabular}{c c}
   \hline
    Representation	& Description   \\ \hline	\hline  \\
$\Patho$  & $\theta$ is constantly in the OFF-state \\ 
& $\Theta(\tau)=0, \forall \tau\in[0,t[$ \\ \\
\hline \\
$\Pathi$  & $\theta$ is switching state at time $s$ \\ 
& $\Theta(\tau)=1$, if $\tau\in [s,t[$ and $\Theta(\tau)=0$ otherwise\\ \\
\hline \\
$\Pathii$
	& $\theta$ is switching state at time $s$ and $s'$\\ 
	& $\Theta(\tau)=1$, if $\tau\in [s,s'[$ and $\Theta(\tau)=0$ otherwise \\ \\
\hline \\
	\end{tabular}
  \caption{Three different types of histories need to be considered. Each path starts with $\theta=0$, and transits no more than once from OFF to ON.}
  \label{table_Path}
\end{table}
\section{Result: final expression of $G^{(0)}(z_1,z_2,t)$}\label{Final expression}
The methodology presented in the previous section leads us to the following generating function (obtained by taking the limit $N\rightarrow\infty$ in equation \eqref{eqGiF}):
\Beq
G^{(0)}(z_1,z_2,t)=
\exp\left[\sum_{j=1,2}(z_j-1)A_j(z_1,z_2,t)\right],
\label{result_0}
\Eeq
where
\Beq
A_j=\int_0^t \mathrm{d}s \int_s^t \mathrm{d}s' \ {\cal K}_j(s,s',t)
\Gii_{\Pathiis},
\Eeq
with the same kernel ${\cal K}_j$ defined in eq. $\eqref{eq_kernel}$).
Finally, $\Gii_\Theta$ for the path $\Pathii$ is explicitly given by
\Beq
\label{Gii_path}
\Gii_{\Pathiis}
		=
\exp
\left[
\sum_{i=1,2}(z_i-1)
\int_s^{s'}\mathrm{d}\lambda \ q_i(\lambda)e^{W_{\gamma_i}(\lambda)-W_{\gamma_i}(t)}
\right].
\Eeq
To keep notation as compact as possible, we will simply write ${\cal K}_j$, omitting the variables $s,s'$ and $t$. The relation between $G^{(0)}$ and $\Gii_\Theta$ is particularly interesting and allows the kernel ${\cal K}_j$ to play a key role in a new set of relations between model-2 and the original model. In model-0, the mean $\moyenne{m_j}$ and correlation $\moyenne{m_j(m_j-1)}$ are obtained using $\moyenne{m_j}(t)=\partial_{z_j}G^{(0)}(z_1,z_2,t)|_{z_1,z_2\rightarrow1}$ and 
$\moyenne{m_j(m_j-1)}(t)=\partial_{z_j}^2G^{(0)}(z_1,z_2,t)|_{z_1,z_2\rightarrow1}$. For the mean, a simple calculation leads to $\moyenne{m_j}(t)=\int_0^t\mathrm{d}s \ \int_s^t\mathrm{d}s' \ {\cal K}_j(s,s',t)$ (identical to eq. \eqref{eqKK}). For the variance, defined by ${\rm{Var}}[m_j](t)=\moyenne{m_j^2}(t)-[\moyenne{m_j}(t)]^2$, writing $\moyenne{m_j}^{(2)}_{\Pathiis}=\partial_{z_j}\Gii_{\Pathiis}|_{z_1,z_2\rightarrow1}$, we obtain :
\Beq\label{VAR_1}
{\rm{Var}}[m_j](t)=\moyenne{m_j}(t)+2\int_0^t\mathrm{d}s \int_s^t\mathrm{d}s' \ {\cal K}_j\moyenne{m_j}^{(2)}_{\Pathiis},
\Eeq
while the correlation function $C_{i,j}(t)=\moyenne{m_im_j}(t)-\moyenne{m_i}(t)\moyenne{m_j}(t)$ becomes
\Beq\label{Cij_1}
C_{i,j}(t)=
\int_0^t\mathrm{d}s \int_s^t\mathrm{d}s' \ \left[{\cal K}_i\moyenne{m_j}^{(2)}_{\Pathiis}(t)+
		{\cal K}_j\moyenne{m_i}^{(2)}_{\Pathiis}(t)\right].
\Eeq
With the mean and correlation numbers in hand, the variance can be {reach}ed easily using ${\rm{Var}}[m_j](t)=\moyenne{m_j}(t)+C_{j,j}(t)$.
It is clear that the generating function can be generalized to an arbitrary number $J$ of proteins by replacing $\sum_{j=1,2}\rightarrow\sum_{j=1}^J$ in equation \eqref{result_0} and \eqref{Gii_path}. In this situation, the generating function depends of $J$ variables: $z_1$, $z_2$, ..., $z_J$.
\subsection{Constant reaction rates}
\begin{figure}
  \centering
  \includegraphics[width=0.5\linewidth]{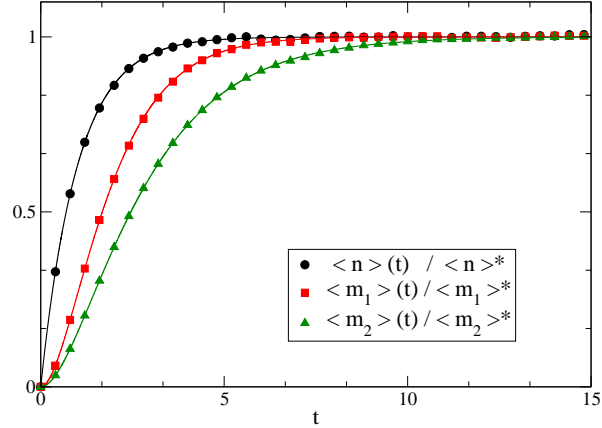}\\
  \caption{\label{xmgrace_fig1} Time evolution of the ratios $\moyenne{n}(t)/\moyenne{n}^*$, $\moyenne{m_1}(t)/\moyenne{m_1}^*$ and $\moyenne{m_2}(t)/\moyenne{m_2}^*$ for the following set of constant parameters: $k=10$, $\mu=1$, $q_1=3$, $\gamma_1=1$, $q_2=5$ and $\gamma_2=1/2$. We observe an excellent agreement between simulation results (circles, squares and triangles) and the analytical expressions (lines). Simulation data, obtained using the Gillespie algorithm, are the result of an average over $10^4$ sampled histories.
} 
\end{figure}

\begin{figure}
  \centering
  \includegraphics[width=0.5\linewidth]{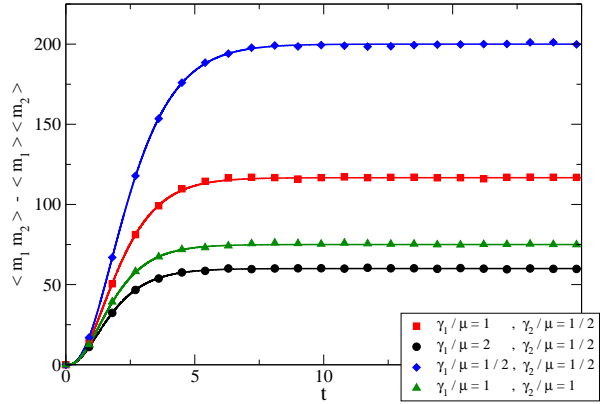}\\
  \caption{    \label{xmgrace_fig3} Time evolution of the correlation function $C(t)=\moyenne{m_1m_2}-\moyenne{m_1}\moyenne{m_2}$ for constant parameter values $k=10$, $\mu=1$, $q_1=3$, $q_2=5$, considering four possible scenarios (1) $(\gamma_1+\gamma_2)/\mu\ne 1$ and $\gamma_j/\mu\ne1$ (black cirlces), (2) $(\gamma_1+\gamma_2)/\mu=1$ (blue diamonds), (3) $\gamma_1/\mu=1$ and $\gamma_2/\mu\ne1$ (red squares) and (4) $\gamma_1/\mu=\gamma_2/\mu=1$ (green triangles). Analytical results (lines) are compared to numerical results (symbols). Simulation data, obtained using the Gillespie algorithm, are the result of an average over $10^5$ sampled histories.} 
\end{figure}

Let us first focus on simplifications occurring when all transition rates are constant. We have $W_\mu(t)=\mu t$, $W_{\gamma_j}(t)=\gamma_j t$ and ${\cal K}_j(s,s',t)=kq_je^{\mu (s-s')+\gamma_j (s'-t)}$. It is then convenient to define $u=e^{\gamma_j(s-t)}$ and $v=e^{\gamma_j(s'-t)}$ so that $A_j$ takes the form
\Beq
A_j=\frac{kq_j}{\gamma_j^2}\int_{e^{-\gamma_jt}}^1\mathrm{d}u \int_u^1\mathrm{d}v \ \Omega_j(u,v),
\Eeq
with 
\Beq
\Omega_j(u,v)=
\frac{u^{\mu/\gamma_j-1}}{v^{\mu/\gamma_j}}\exp\left[{\sum_{i}(z_i-1)\frac{q_i}{\gamma_i}(v^{\gamma_i/\gamma_j}-u^{\gamma_i/\gamma_j})}\right].
\Eeq
At this stage it is not hard to show that the mean protein number is given by \eqref{eq5}. Figure (\ref{xmgrace_fig1}) confirms the validity of our results. For correlation numbers, it is useful to define the normalised function $\tilde C_{i,j}(t)=C_{i,j}(t)/{\moyenne{m_i}^*\moyenne{m_j}^*}$.
For $\gamma_i+\gamma_j\ne\mu$, $\gamma_i\ne\mu$ and $\gamma_j\ne\mu$ the latter quantity is given by
\Beq
\label{C_1}
\frac{k}{\mu} \tilde C_{i,j}&=&
\frac{\gamma_i+\gamma_j}{\gamma_i+\gamma_j-\mu}\left(1-e^{-\mu t}\right) \\
&-&
\frac{\gamma_i}{\gamma_i-\mu}\frac{\mu}{\gamma_j+\mu}
\left[1-e^{-(\gamma_j+\mu)t}\right]
-
\frac{\gamma_j}{\gamma_j-\mu}\frac{\mu}{\gamma_i+\mu}
\left[1-e^{-(\gamma_i+\mu)t}\right]\nonumber \\
&+&
\frac{\mu}{\gamma_i+\gamma_j}
\frac{\gamma_i}{\gamma_i-\mu}
\frac{\gamma_j}{\gamma_j-\mu}
\frac{\gamma_i+\gamma_j-2\mu}{\gamma_i+\gamma_j-\mu}
\left[1-e^{-(\gamma_i+\gamma_j)t}\right] . \nonumber
\Eeq
The case $\gamma_i=\mu$ (or $\gamma_j=\mu$) has to be treated separately. To proceed one can (1) set $\gamma_i=\mu$ in the kernel $K_i$, or alternatively (2) write $\gamma_i=\mu+\epsilon$ and take the limit $\epsilon\rightarrow 0$. The case $\gamma_i+\gamma_j=\mu$  (or $2\gamma_i=\mu$ when considering the variance) has to be treated similarly. Those limits lead to relatively more compact expressions, for example when $\gamma_i=\gamma_j=\mu$, we have:
\Beq
\frac{k}{\mu}&\tilde C_{i,j} &=\frac{1}{2}-2e^{-\mu t}+e^{-2\mu t}(3/2+\mu t).
\Eeq
The agreement (for all time $t$) between analytical expressions and numerical simulations can be seen in Figure (\ref{xmgrace_fig3}). In the limit $t\rightarrow\infty$ all expressions of $\tilde C_{i,j}$ converge to a single form. In the stationary state, the correlation function has a unique expression
\Beq
\tilde C^*_{i,j}=\frac{\mu}{k}\frac{\eta_i}{\eta_i+1}\frac{\eta_j}{\eta_j+1}
\frac{\eta_i+\eta_j+2}{\eta_i+\eta_j},
\label{renormalized_Corr_result_stationary}
\Eeq
with $\eta_i=\gamma_i/\mu$ and $\eta_j=\gamma_j/\mu$. So that
\Beq\label{eq62C}
C^*_{i,j}=\frac{k}{\mu}\frac{q_i}{\gamma_i+\mu}\frac{q_j}{\gamma_j+\mu}
\frac{\gamma_i+\gamma_j+2\mu}{\gamma_i+\gamma_j}.
\label{Corr_result_stationary}
\Eeq
For homogeneous degradation rates ($\gamma_i=\gamma_j=\gamma$), the correlation is invariant under the exchange $\gamma\leftrightarrow\mu$. The last equation clearly shows that the correlation function does not vanish (unless one out of $\moyenne{m_i}^*$ and $\moyenne{m_j}^*$ vanishes). We note that the correlation $C^*_{i,j}$ is strictly monotonic (decreasing) in terms of $\gamma_i$ and $\gamma_j$ (keeping all other parameters constant). As a consequence, if one can estimate lower and upper bounds of both $\gamma_i$ and $\gamma_j$ it is, in principle, possible to restrain the range of correlation values to an interval: $[C^*_{min},C^*_{max}]$. In addition, we observe, for a fixed value of $\eta_j$, that the correlation $\tilde C^*$ presents a maximum at $(\eta_i)_{max}=\eta_j+\sqrt{2\eta_j(\eta_j+1)}$. If $C^*$ is strictly monotonic, it is when varying $\gamma_i$ while keeping $\moyenne{m_i}^*$ constant that a non monotonic behaviour is observed. In this case, $C^*_{i,j}$ can be rewritten as
\Beq
C^*_{i,j}=\moyenne{m_i}^*\frac{q_j/\mu}{\eta_j+1}
\frac{\eta_i}{\eta_i+1}\frac{\eta_i+\eta_j+2}{\eta_i+\eta_j},
\Eeq
and presents a maximum in $(\eta_i)_{max}$. Keeping both protein levels $\moyenne{m_i}^*$ and $\moyenne{m_j}^*$ constant, the correlation function becomes $C^*_{i,j}=\moyenne{m_i}^*\moyenne{m_j}^*\tilde C^*_{i,j}$. Looking for an upper bound into the $(\eta_i,\eta_j)$-plane, one needs to solve $\partial_{\eta_i}C^*=0$ and $\partial_{\eta_j}C^*=0$ simultaneously. However there are no strictly positive solutions to the latter system of equations. Hence, under this constrain, $C^*_{i,j}$ does not present a maximum when varying both $\gamma_i$ and $\gamma_j$.

\subsubsection{The $2$-stage model: $J=1$}
In the case $J=1$, the model with constant transition rates, reduces to the conventional two-stage model. To pursue, we define $r=\mu/\gamma$ and $\delta(z)=q(z-1)/\gamma$. We can show that our result leads to the solution first presented in \cite{Bokes_2012} and later in \cite{Pendar_2013}:
\Beq
\label{past_solution}
G^*(z)=\exp\left(\frac{k}{\mu}\int_0^{\delta(z)} \mathrm{d}s \ _1F_1\left[1,r+1,s\right]\right),
\Eeq
where $\hphantom{.}_1F_1$ is the confluent hypergeometric function. As it is, the identity between equation \eqref{result_0} (for one protein type only) and equation \eqref{past_solution} is not obvious. To proceed, we use the Taylor expansion of $e^{\delta v}$ and $e^{-\delta u}$ and write $\epsilon=e^{-\gamma t}$ which we assume small compared to one. Considering $r=\mu/\gamma\ne1$ and keeping the lowest order in $\epsilon$ (see appendix) we show that 
\Beq\label{approx}
G^{(0)}(z,t)\underset{t\gg1}{\simeq}G^*(z)H(z,t),
\Eeq 
with $G^*(z)$ given by equation \eqref{past_solution} and 
\begin{equation}
\label{eq_HH}
\ln(H(z,t))
=\left\{
\begin{array}{cc}
\frac{k}{\gamma-\mu}e^{-\gamma t}\delta(z)& \gamma<\mu\\
-\frac{k}{\mu}e^{-\mu t}\sum_{m=0}^\infty \frac{(\delta(z))^{m+1}}{m!({m+1-r})}  & \gamma>\mu,
\end{array}
\right.
\end{equation}
such that $\lim_{z\rightarrow1} H(z,t)=\lim_{t\rightarrow\infty} H(z,t)=1$. The latter approximation leads to:
\begin{equation}
\frac{\moyenne{m}(t)}{\moyenne{m}^*}\underset{t\gg1}{\simeq}1
+
\frac{1}{\gamma-\mu}
\left\{
\begin{array}{cc}
\hphantom{(+1)}{\mu}e^{-\gamma t}& \gamma<\mu\\
(-1)\gamma e^{-\mu t}
  & \gamma>\mu,
\end{array}
\right.
\end{equation}
in agreement with equation \eqref{eq5}. The case $\gamma=\mu$ is treated separately in appendix.
\subsubsection{Homogeneous degradation rates: $\gamma_j=\gamma$, $\forall j$}
When dealing with $J$ protein types ($J>1$) and homogeneous degradation rates ($\gamma_j=\gamma$ $\forall j$), the generating function reduces to a form close to the one previously obtained for the two-stage model. Defining $\Delta(\{z_j\})=\sum_jq_j(z_j-1)/\gamma$, we can show that $
G^{(0)}(\{z_j\},t)$ is given by equation \eqref{approx} under the substitution $\delta(\{z_j\})\rightarrow\Delta(\{z_j\})$. It follows that
\Beq
\lim_{t\rightarrow\infty}G^{(0)}(\{z_j\},t)=
\exp\left(\frac{k}{\mu}\int_0^{\Delta(\{z_j\})} \mathrm{d}s \ _1F_1\left[1,r+1,s\right]\right).
\Eeq
The generating function ${\cal G}$, associated to the total number of proteins ($M=\sum_jm_j$), is defined by ${\cal G}(z)=\sum_MP_Mz^M$, with
\Beq
P_M=\sum_{m_1,m_2,\hdots,m_J}P_{m_1,m_2,\hdots,m_J}\delta\left(\sum_jm_j-M\right).
\Eeq
We see that ${\cal G}$ is given by ${\cal G}(z,t)=G^{(0)}(\{z_j=z\},t)$:
\Beq
{\cal G}^*(z)=
\exp\left(\frac{k}{\mu}\int_0^{J\bar q(z-1)/\gamma} \mathrm{d}s \ _1F_1\left[1,r+1,s\right]\right),
\Eeq
with the average creation rate defined by $J \bar q=\sum_j q_j$. The mean of total protein number ($M=\sum_j m_j$) satisfies $
{\moyenne{M}}/{J}={k}{\bar q}/({\mu}{\gamma})$.

\subsection{Time dependent transition rates: a bridge towards other models}
\begin{figure}
  \centering
  \includegraphics[width=0.6\linewidth]{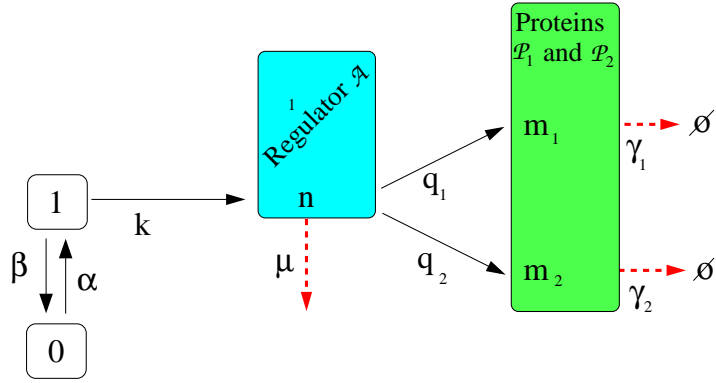}\\
  \caption{\label{FIG_3_Stage} The three-stage model of gene expression, with arbitrary {partition} of proteins.} 
\end{figure}

\begin{figure}
  \centering
  \includegraphics[width=0.65\linewidth]{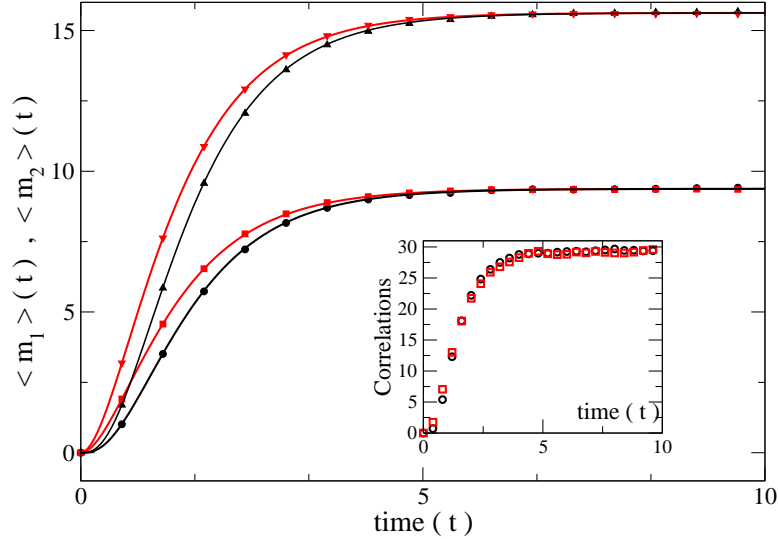}\\
  \caption{    \label{xmgrace_fig4} Comparison of the time evolution of the mean numbers $\moyenne{m_1}(t)$,  $\moyenne{m_2}(t)$ for the two stage model (in red - for parameter values $k=6.25$, $\mu=1$, $q_1=3$, $q_2=5$, $\gamma_1=\gamma_2=2$) and three state model (in black - for parameter values $\alpha=2.5$, $\beta=1.5$, $k_1=10$, $\mu=1$, $q_1=3$, $q_2=5$, $\gamma_1=\gamma_2=2$). Analytical results (lines) are compared to numerical results (symbols). In the insert, we compare correlations between protein numbers for the two stage model (in red) and three stage model (in black). For the same parameter values, we plot the time evolution of $(1+R)C_{1,2}(t)$ (red) together with $\CC_{1,2}(t)$ (black).} 
\end{figure}

Results for time dependent parameters {allow} for the study of fluctuations (induced by hidden variables) in production and/or degradation rates. In a recent paper Dattani and Barahona \cite{Dattani_2017} proposed a framework to model gene expression with stochastic or deterministic transcription and degradation rates. Along the same lines, let us start by defining random variables $x_\phi$, for all parameters $\phi$ of the model ($\phi=k,\mu,q_1,q_2,...,\gamma_1,\gamma_2,...$). We choose to write $\phi(t)=\phi_0+\phi_1x_\phi(t)$ with $\phi_0,\phi_1\in{\mathbb R}$, but other functional forms could be chosen. More explicitly, we have for $k(t)$: $k(t)=k_0+k_1x_k(t)$. A particular time history of the random variable $x_k(t)$ is written $X_k=(x_k(t)|\forall t)$. Identically, we write $X_\phi=(x_\phi(t)|\forall t)$ the history for random variable $x_\phi$. Finally, we define ${\mathbb X}$ as the set of histories ${\mathbb X}=\{X_k,X_\mu,X_{q_1},...X_{\gamma_1},...\}$ so that the generating function is now explicitly dependent on ${\mathbb X}$: we write $G^{(0)}_{{\mathbb X}}$. Writing $\PP({\mathbb X})$ as the probability of the set of histories, the new generating function is 
\Beq
\moyenne{G^{(0)}}=\sum_{{\mathbb X}}\PP({\mathbb X})G^{(0)}_{{\mathbb X}},
\Eeq
where $\sum_{{\mathbb X}}$ symbolically represents the sum over all possible histories of all parameters. The mean protein number is given by
\Beq
\moyenne{\moyenne{m_j}}=\sum_{{\mathbb X}}\PP({\mathbb X})\moyenne{m_j}_{{\mathbb X}},
\Eeq
with $\moyenne{m_j}_{{\mathbb X}}=\partial_{z_j}G^{(0)}_{{\mathbb X}}|_{\forall z_i=1}$. 
To give a concrete example, we will restrict ourself to time dependent production rate $k(t)$ while all other transition rates are constant. We show how our results bridge towards the three-stage model, {allow}ing us to access the exact mean and correlation functions. In this situation ${\cal K}_j(s,s',t)=k(s)q_je^{\mu (s-s')+\gamma_j (s'-t)}$. Choosing the appropriate function $k(t)$ can give information on the behaviour induced by state fluctuation of the DNA operational site (Figure \eqref{FIG_3_Stage}). We will write $\moyenne{m_j}_{X_k}$ the mean protein numbers, for a particular history ${X_k}$. For $\gamma_j\ne\mu$ and without restriction on $k(t)$, we show that
\Beq\label{convo_m}
\moyenne{m_j}_{X_k}(t)=\frac{q_j}{\gamma_j-\mu}k\star\left[e^-_{\mu}- e^-_{\gamma_j}\right],
\Eeq
with the convolution product
\Beq
(k\star e^\pm_{a})(t)=\int_0^t\mathrm{d}s \ k(s)e^{\pm a(t-s)}.
\Eeq
For $\gamma_j=\mu$ we write $\gamma_j=\mu+\epsilon$ in \eqref{convo_m} together with the limit $\epsilon\rightarrow 0$. \footnote{Along the same line, when considering time dependent production rate $q_j(t)$ while keeping all other parameters constant, we get an equation similar to (\ref{convo_m}):
$
\moyenne{m_j}_{X_{q_j}}(t)=({k}/{\mu})\left[q_j-q_je^-_{\mu}\right]\star e^-_{\gamma_j}.
$
}
The previous equation becomes $
\moyenne{m_j}_{X_k}(t)=q_j\left(-\frac{\partial}{\partial\mu}\right)(k\star e^-_{\mu})(t)$. For all values of $\gamma_j$ and $\mu$, the Laplace transform of the mean number of protein $L[\moyenne{m_j}]$ simplifies to a single expression:
\Beq
L[\moyenne{m_j}_{X_k}](s)=\frac{q_j L[k](s)}{(s+\mu)(s+\gamma_j)},
\Eeq
with $L[k]$ as the Laplace transform of $k(t)$. Assuming the limit $k(t\rightarrow\infty)=k^*$ exists, the final value theorem leads to $\moyenne{m_j}^*=k^*q_j/(\gamma_j\mu)$. With the three-stage model in mind, we set $k_0=0$ and write $k(t)=k_1x_k(t)$, where the random variable $x_k(t)$ takes value in $\{0,1\}$. It describes the possible states, active ($x_k =1$) or inactive ($x_k=0$), of the promoter region. Governed by a simple two state dynamics (with transition rates $W_{0\rightarrow1}=\alpha$ and $W_{1\rightarrow0}=\beta$) the variable $x_k(t)$ ``oscillates'' between those states (see illustration \eqref{FIG_3_Stage}). This motivates in \cite{Jedrak_2016} the choice of a sinusoidal function: $k(t)=c_1\sin(\omega t+\phi)+c_2$. However, the time evolution of the variable $x_k(t)$ is stochastic and, starting from initial condition $\moyenne{x_k}(t=0)=0$, it satisfies
\Beq
\frac{\moyenne{x_k}(t)}{\moyenne{x_k}^*}=\chi(t)=1-e^{-(\alpha+\beta)t},
\Eeq
with stationnary state $\moyenne{x_k}^*=\alpha/(\alpha+\beta)$. It follows that the mean for the three stage model is given by $\moyenne{\moyenne{m_j}}$ representing the average over the history of the variable $x_k$. For $\gamma_j\ne\mu$ we have to evaluate
\Beq
\frac{\moyenne{\moyenne{m_j}}(t)}{\moyenne{\moyenne{m_j}}^*}
=
\frac{\mu\gamma_j}{\gamma_j-\mu}
\chi\star\left[e^-_{\mu}- e^-_{\gamma_j}\right]
\Eeq
with $\moyenne{\moyenne{m_j}}^*=\moyenne{x_k}^* k_1q_j/(\mu\gamma_j)$. A simple calculation leads to the exact expression:
\Beq\label{equation_mean_3_states}
\frac{\moyenne{\moyenne{m_j}}(t)}{\moyenne{\moyenne{m_j}}^*}=1
&+&\frac{\mu}{\gamma_j-\mu}\frac{\alpha+\beta}{\alpha+\beta-\gamma_j}e^{-\gamma_j t}
+\frac{\gamma_j}{\mu-\gamma_j}\frac{\alpha+\beta}{\alpha+\beta-\mu}e^{-\mu t}\nonumber\\
&-&\frac{\mu}{\alpha+\beta-\mu}\frac{\gamma_j}{\alpha+\beta-\gamma_j}e^{-(\alpha+\beta) t},
\Eeq
as long as $\gamma_j\ne \mu$, $\mu\ne\alpha+\beta$ and $\gamma_j\ne\alpha+\beta$.  Once again, the Laplace transform gives one single expression valid in all parameter space:
\Beq \label{Laplace_equation_mean_3_states}
L\left[\frac{\moyenne{\moyenne{m_j}}}{\moyenne{\moyenne{m_j}}^*}\right](s)=
\frac{1}{s}
\frac{\alpha+\beta}{s+\alpha+\beta}
\frac{\mu}{s+\mu}
\frac{\gamma_j}{s+\gamma_j}.
\Eeq
This result is not new and could have alternatively been obtained by writing $d\moyenne{\moyenne{m_j}}/dt=q_j\moyenne{\moyenne{n}}-\gamma_j\moyenne{\moyenne{m_j}}$, which solution is $\moyenne{\moyenne{m_j}}=q\int_0^t\mathrm{d}s \ \moyenne{\moyenne{n}}(s)e^{-\gamma_j(t-s)}$ and using the time evolution of mRNA level $\moyenne{\moyenne{n}}$ (presented in \cite{Peccoud_1995}). Figure (\ref{xmgrace_fig4}) shows agreement between analytical predictions and numerical simulations. If it is mathematically convenient to consider the time evolution starting from an ``empty'' initial state (all stochastic variable to zero), this situation does not seem to be biologically relevant. One could however, consider the similar scenario starting from the state $x_k=0$, with initial numbers $n,m_1,m_2$ of $A$ macromolecules and proteins. Solving this new problem requires a different approach based on an variation of the PPA mapping, which is not considered in this paper.

Finally, let us discuss how to infer on correlation numbers between protein types. First, with the help of 
\Beq
\moyenne{m_j}^{(2)}_{\Pathiis}(t)=\frac{q_j}{\gamma_j}\left(e^{\gamma_j(s'-t)}-e^{\gamma_j(s-t)}\right),
\Eeq
we can use equation \eqref{Cij_1} to express the correlation function $C_{i,j|{X_k}}=\moyenne{m_im_j}_{X_k}(t)-\moyenne{m_i}_{X_k}(t)\moyenne{m_j}_{X_k}(t)$:
\Beq\label{convo_c}
\frac{C_{i,j|{X_k}}}{(q_i/\gamma_i)(q_j/\gamma_j)}&=&k\star\left[
\frac{\gamma_i+\gamma_j}{\gamma_i+\gamma_j-\mu}e^-_\mu \right. 
-\frac{\gamma_i}{\gamma_i-\mu}e^-_{\gamma_j+\mu}
-\frac{\gamma_j}{\gamma_j-\mu}e^-_{\gamma_i+\mu}\nonumber\\
&+&
\left. \frac{\gamma_i}{\gamma_i-\mu}\frac{\gamma_j}{\gamma_j-\mu}
\frac{\gamma_i+\gamma_j-2\mu}{\gamma_i+\gamma_j-\mu}e^-_{\gamma_i+\gamma_j}
\right].
\Eeq
One should note that the variance is given by $C_{j,j|{X_k}}(t)+\moyenne{m_j}_{X_k}(t)$ and can be evaluated using equations \eqref{convo_m} and \eqref{convo_c}. For singular cases $\gamma_i=\mu$, $\gamma_j=\mu$ or $\gamma_i+\gamma_j=\mu$ a similar expression can be derived from the previous equation taking the limit appropriately. To continue further one has to proceed more carefully. In fact the correlations in the model presented in figure 6 are defined by $\CC_{i,j}=\moyenne{\moyenne{m_im_j}}-\moyenne{\moyenne{m_i}}\moyenne{\moyenne{m_j}}$, which can be expressed using $\moyenne{C_{i,j}}$ (the average of $C_{i,j|X_k}$ over the history $X_k$): 
\Beq\label{eq_CIIJJ}
\CC_{i,j}=\moyenne{C_{i,j}}+\moyenne{\moyenne{m_i}\moyenne{m_j}}-\moyenne{\moyenne{m_i}}\moyenne{\moyenne{m_j}}.
\Eeq
One can notice that $\moyenne{C_{i,j}}$ (just like $\moyenne{\moyenne{m_i}}$ and $\moyenne{\moyenne{m_j}}$) is a functional of the mean $\moyenne{x_k}(t)$, and can be evaluated easily. The challenge comes from the term $\moyenne{\moyenne{m_i}\moyenne{m_j}}$ as it requires knowledge of correlators $\moyenne{x_k(s)x_k(s')}$:
\Beq
& &\moyenne{\moyenne{m_i}\moyenne{m_j}}=
\frac{k_1^2q_iq_j}{(\gamma_i-\mu)(\gamma_j-\mu)}\times\\ \nonumber
& &\int_0^t\mathrm{d}s \int_0^t\mathrm{d}s'\
\moyenne{x_k(s)x_k(s')}(e^{-\mu(t-s)}-e^{-\gamma_i(t-s)})
(e^{-\mu(t-s')}-e^{-\gamma_j(t-s')}).
\Eeq
In fact the $x_k$-correlation can be calculated exactly. For $s<s'$, it is given by 
\Beq
\moyenne{x_k(s)x_k(s')}
=
\left(\frac{\alpha}{\alpha+\beta}\right)^2
\left(1-e^{-(\alpha+\beta)s}\right)
\left(1+\frac{\beta}{\alpha}e^{-(\alpha+\beta)(s'-s)}\right).
\Eeq
This last expression, together with the help of \eqref{convo_c} and \eqref{equation_mean_3_states} allow for the evaluation of ${\cal C}_{i,j}$. To compare correlations $\CC_{i,j}$ for the three-stage model with $C_{i,j}$ \eqref{eq62C} for the two-stage model, we impose the equality $\alpha k_1=(\alpha+\beta)k$ which assures, in both models, identical regulator, and proteins levels. It follows that ${\CC_{i,j}^*}/{C_{i,j}^*}=1+R$, where $R$ has a cumbersome expression, dependent on all parameters but $q_i$ and $q_j$. It vanishes for $\beta=0$ (as well as $k=0$) and satisfies $R>0$ for all other finite parameter values. Hence we conclude that, in the stationary state, for identical regulator and protein levels, correlations between protein numbers are higher in the model with promoter-based regulation: $\CC_{i,j}^*>C_{i,j}^*$. Restraining ourself to homogeneous degradation rates ($\gamma_i=\gamma_j=\gamma$) the expression for $R$ is more manageable :
\Beq 
R=\frac{\beta}{\alpha+\beta}\frac{k_1(\alpha+\beta+\mu+\gamma)}{(\alpha+\beta+\mu)(\alpha+\beta+\gamma)}.
\Eeq
Once again, we note that $R$ and $\CC_{i,j}^*$ are invariant under the exchange $\mu\leftrightarrow\gamma$. Figure (\ref{xmgrace_fig4}) shows a comparison between the time evolution of protein number in the two and three stage model. Our data validate the equality ${\CC_{i,j}^*}/{C_{i,j}^*}=1+R$ and seem to indicate that $\CC_{i,j}(t)/C_{i,j}(t)\simeq 1+R$ is an acceptable approximation, at least for the set of parameter selected.

\section{Conclusion}
In this paper, we present a variation on the two-stage model including the arbitrary {partition} of protein and arbitrary time dependent parameters. The mechanism considered is similar to the one involved in alternative splicing \cite{Wang_2014}. Our aim was to develop an analytical approach leading to the time dependent generating function, based on the PPA-mapping, and which did not require knowledge of already known results. Constructed on a succession of transformations, our work simplifies the original problem to the analysis of two-states biological switches. As a consequence, a series of different reduced models emerges, with clear relations linking their respective generating functions. In particular we show how the kernel ${\cal K}_j$ plays a important role in the final expression of the generating function. We show that the emerging hierarchy {allow}s us to connect mean numbers and higher order moments between models. This leads to an explicit relation between correlation $C_{i,j}$ in the original model and mean protein number in model-2. For constant parameter values, we derived the exact time dependent expression of correlation $C_{i,j}$. We note that particular cases such as $\gamma_j=\mu$, $\gamma_j=\mu/2$ or $\gamma_1+\gamma_2=\mu$ have to be treated separately. However, in the stationary state this distinction vanishes as each function converges towards a common asymptotic expression. Considering constant parameters, for $J=1$ or for homogeneous degradation rates ($\gamma_j=\gamma$ $\forall j$), we show how our results reduce to the solution of two stage model. Finally, we show how results for arbitrary time dependent transition rates can be used to study models presenting parameter fluctuations induced by hidden random variable. We give an explicit example by extending our results for the mean and correlation to the three state model. In particular we show that promoter based regulation leads to higher protein-protein correlations $\CC^*_{i,j}>C_{i,j}^*$. The method presented here is applicable in other scenarios (with zero and first order reactions) and may be use as a guide in the study of other biological systems. We hope this methodology will contribute to the development of new analytical avenues for future research.

\section*{Acknowledgements}
The authors would like to thanks the AMRC in Coventry and the Stat. Phys. Group in Nancy for their support and constant efforts to protect curiosity driven research. In particular, T. Platini extends his acknowledgements to K. Aujogue, R. Low, S. Mukherjee, S. Vantieghem and R. Kulkarni.

\section{Appendix}
\subsection{From model-0 to model-1: a detailed derivation}
In the following we aim to construct the original problem (model-0) starting with $N$ independent models, all identical to model-1. The derivation is presented for constant parameter, but could easily be extended to time dependent parameter. We will start with the definition of $\phi_{N;(\theta,a,b)}(t)$: the probability of finding, in model-1, the biological switch in the state $\theta$ together with $a$ and $b$ proteins of type $\PP_1$ and $\PP_2$ respectively. We continue by defining $\Phi_N(n,m_1,m_2)$ as the probability of finding amongst $N$ independent identical models (each labeled with subscript $\nu$) the total numbers of active switches $n=\sum_\nu \theta_\nu$ and proteins $m_1=\sum_\nu a_\nu$, $m_2=\sum_\nu b_\nu$. Finally when, taking the limit $N\rightarrow\infty$, we will be able to show that $\lim_{N\rightarrow\infty}\Phi_{N}(t)$ obeys the master equation $\eqref{eq_Master_eq}$. In other words $P_{n,m_1,m_2}=\lim_{N\rightarrow\infty}\Phi_{N;(n,m_1,m_2)}$.

To start let us try to keep notation as compact as possible by defining the flip operator $\hatf$ as the operator acting on the triplet $(\theta,a,b)$ and such that $\hatf(\theta,a,b)=(1-\theta,a,b)$. In addition, we write $\hata^\pm$ and $\hatb^\pm$ as the operators defined by
\Beq
\hata^\pm(\theta,a,b)=(\theta,a\pm1,b),\\
\hatb^\pm(\theta,a,b)=(\theta,a,b\pm1).
\Eeq
Let $\phi_{N;(\theta,a,b)}(t)$ be the probability distribution associated to model-1. It obeys the following equation
\Beq
\frac{d \phi_{N;(\theta,a,b)}}{dt} &=&
\left[\theta(k/N)+(1-\theta)\mu \right]\phi_{N;\hatf(\theta,a,b)}\\
&+& \theta q_1 \phi_{N;\hata^-(\theta,a,b)} + \theta q_2 \phi_{N;\hatb^-(\theta,a,b)} \nonumber \\
&+& (a+1)\gamma_1 \phi_{N;\hata^+(\theta,a,b)} + (b+1)\gamma_2 \phi_{N;\hatb^+(\theta,a,b)}\nonumber\\
&-& \left[ (1-\theta)(k/N)+\theta\mu+\theta q_1+\theta q_2+a\gamma_1+b\gamma_2 \right]
\phi_{N;(\theta,a,b)}
.\nonumber
\Eeq
Considering $N$ independent but identical models, we define $\Sbb$ as the set of triplets: $\Sbb=\{(\theta_\nu,a_\nu,b_\nu);\nu=1,2,\hdots, N\}$. Note that $\Sbb$ gives a full description of the state of all $N$ independent reduced models. In addition, we will write $\theta_\nu(\Sbb)$, $a_\nu(\Sbb)$ and $b_\nu(\Sbb)$ the variable $\theta$, $a$ and $b$ in the $\nu^{th}$ triplet of the set $\Sbb$. The definitions of the operator $\hatf$, $\hata^{\pm}$ and $\hatb^{\pm}$ are extended onto the set $\Sbb$. So that $\hatf_\nu$, $\hata^\pm_\nu$ and $\hatb^\pm_\nu$ act on the $\nu^{th}$ triplet of the set $\Sbb$, leaving all others unchanged. We can now define the overall probability $\phi_{N,\Sbb} = \prod_{\nu=1}^N \phi_{N,(\theta_\nu,a_\nu,b_\nu)}$,
which evolution is governed by
\Beq
& &\frac{d}{dt}\phi_{N,\Sbb} = \sum_{i=1}^N\prod_{\nu\ne i} \phi_{N,(\theta_\nu,a_\nu,b_\nu)} 
\frac{d}{dt}\phi_{N,(\theta_i,a_i,b_i)}.
\Eeq
With a little bit of effort, the latter equation leads to
\Beq
\frac{d}{dt} \phi_{N,\Sbb} &=&
k / N \sum_{i}
\left\{
\theta_i(\Sbb)\phi_{N,\hatf_i\Sbb} 
-
[1-\theta_i(\Sbb)] \phi_{N,\Sbb}
\right\} 
\nonumber \\
& + &
\mu\sum_{i}
\left\{
[1-\theta_i(\Sbb)]\phi_{N,\hatf_i\Sbb} 
-
\theta_i(\Sbb) \phi_{N,\Sbb}
\right\}
\nonumber \\
& + & 
q_1 \sum_{i} 
\theta_i(\Sbb) \left\{
 \phi_{N,\hata^-_i\Sbb} 
- 
 \phi_{N,\Sbb} 
\right\}
 \\
& + & q_2 \sum_{i} 
\theta_i(\Sbb) \left\{
 \phi_{N,\hatb^-_i\Sbb} 
-
\phi_{N,\Sbb}
\right\}
  \nonumber \\
&+& \gamma_1 \sum_{i}
\left\{ 
[a_i(\Sbb)+1] \phi_{N,\hata^+_i\Sbb} 
-
a_i(\Sbb) \phi_{N,\Sbb}
\right\}
  \nonumber \\
&+& \gamma_2 \sum_{i} 
\left\{
[b_i(\Sbb)+1] \phi_{N,\hatb^+_i\Sbb}
-
b_i(\Sbb) \phi_{N,\Sbb}
\right\}.
 \nonumber
\Eeq
Moving forward, we define the probability distribution $\Phi_N$ as
\Beq
\Phi_{N;(n,m_1,m_2)}=\sum_{\Sbb}\phi_{N;\Sbb}
\Delta^{\{\theta_\nu\}}_n
\Delta^{\{a_\nu\}}_{m_1}
\Delta^{\{b_\nu\}}_{m_2},
\Eeq
where the constraint $\Delta$ is defined by
\Beq
\Delta^{\{x_\nu\}}_{y}=\delta\left(\sum_\nu x_\nu,y\right),
\Eeq
where $\delta$ is the Kronecker symbol. To keep notations as compact as possible we write $\Delta^{(3)}=\Delta^{\{\theta_\nu\}}_n\Delta^{\{a_\nu\}}_{m_1}\Delta^{\{b_\nu\}}_{m_2}$. The master equation for the distribution $\Phi_N$ is
\Beq
\frac{d}{dt}\Phi_{N,(n,m_1,m_2)} &=&
k/N
\sum_{\Sbb}
\Delta^{(3)}
 \sum_{i}
 \left\{
 \theta_i(\Sbb)\phi_{N,\hatf_i\Sbb} 
 -
 [1-\theta_i(\Sbb)] \phi_{N,\Sbb}
 \right\}
 \\ \nonumber 
&+&
\mu
\sum_{\Sbb}
\Delta^{(3)}
\sum_{i}
\left\{
[1-\theta_i(\Sbb)]\phi_{N,\hatf_i\Sbb} 
-
\theta_i(\Sbb) \phi_{N,\Sbb}
\right\}
\nonumber \\
&+& 
q_1
\sum_{\Sbb}
\Delta^{(3)}
\sum_{i} 
\left\{
\theta_i(\Sbb) \phi_{N,\hata^-_i\Sbb} 
-
\theta_i(\Sbb)  \phi_{N,\Sbb} 
\right\}
 \nonumber\\
&+&
q_2 
\sum_{\Sbb}
\Delta^{(3)}
\sum_{i} 
\left\{
\theta_i(\Sbb) \phi_{N,\hatb^-_i\Sbb} 
-
\theta_i(\Sbb)  \phi_{N,\Sbb}
\right\}
  \nonumber \\
&+&
\gamma_1
\sum_{\Sbb}
\Delta^{(3)}
  \sum_{i} 
  \left\{
  [a_i(\Sbb)+1] \phi_{N,\hata^+_i\Sbb} 
 -
a_i(\Sbb) \phi_{N,\Sbb} 
  \right\}
 \nonumber\\
&+& 
\gamma_2
\sum_{\Sbb}
\Delta^{(3)}
\sum_{i} 
\left\{
[b_i(\Sbb)+1] \phi_{N,\hatb^+_i\Sbb}
-
b_i(\Sbb) \phi_{N,\Sbb}
\right\}.
 \nonumber
\Eeq
Every sum, for which $\phi_{N;\Sbb}$ appears explicitly can be easily evaluated. As an example we give here the second term of the first line in the previous equation. Using the {constraint} $\Delta^{\{\theta_\nu\}}$, which impose $\sum_\nu\theta_\nu=n$, we have
\Beq
\sum_{\Sbb}
\Delta^{\{\theta_\nu\}}_n
\Delta^{\{a_\nu\}}_{m_1}
\Delta^{\{b_\nu\}}_{m_2} 
\sum_{i} [1-\theta_i(\Sbb)] \phi_{N,\Sbb}
&=&
\sum_{\Sbb}
\Delta^{\{\theta_\nu\}}_n
\Delta^{\{a_\nu\}}_{m_1}
\Delta^{\{b_\nu\}}_{m_2} 
[N-n] \phi_{N,\Sbb}\nonumber
\\
&=&
(N-n)\Phi_{N,(n,m_1,m_2)}.
\Eeq
When $\phi_{N,\Sbb}$ does not appear explicitly we need to re-labeled the sum over $\Sbb$. As an example we present details to the calculation of the first term of the first line in which we have $\phi_{N,\hatf_i\Sbb}$. Defining $\tilde\Sbb=\hatf_i\Sbb$ we have $\hatf_i\tilde\Sbb=\Sbb$ so that 
\begin{equation}
\theta_j(\Sbb)=\theta_j(\hatf_i\tilde\Sbb)=\left\{
\begin{array}{ccc}
\theta_\nu(\tilde\Sbb), &{\text{if}}& i\ne \nu\\
1-\theta_i(\tilde\Sbb), &{\text{if}}& i=\nu.
\end{array}
\right.
\end{equation}
In relabelling $\Sbb$ to $\tilde\Sbb$ the expression of $\Delta^{\{\theta_\nu\}}$ has changed. To keep track of this change we will replace $\Delta^{\{\theta_\nu\}}\rightarrow\tilde\Delta^{\{\theta_\nu\}}$ where
\Beq
\Delta^{\{\theta_\nu\}}_n&=&\delta\left(\sum_\nu\theta_\nu(\Sbb),n\right)=\delta\left(\sum_{\nu}\theta_\nu(\tilde\Sbb)+1-2\theta_i(\tilde\Sbb),n\right)
=\tilde\Delta^{\{\theta_\nu\}}_n
\Eeq
It follows that 
\Beq
\sum_{i=1}^N\sum_{\Sbb}
\Delta^{\{\theta_\nu\}}_n
\Delta^{\{a_\nu\}}_{m_1}
\Delta^{\{b_\nu\}}_{m_2}
\theta_i(\Sbb)\phi_{N,\hatf_i\Sbb} 
=
\sum_{i=1}^N\sum_{\tilde\Sbb}
\tilde\Delta^{\{\theta_\nu\}}_n
\Delta^{\{a_\nu\}}_{m_1}
\Delta^{\{b_\nu\}}_{m_2}
[1-\theta_i(\tilde\Sbb)]\phi_{N,\tilde\Sbb} .\nonumber\\
\Eeq
Note that the only elements which will contribute are such that $\theta_i(\tilde\Sbb)=0$, which {allow}s us to write 
\Beq
\tilde\Delta^{\{\theta_j\}}_n
=\delta\left(\sum_{\nu}\theta_\nu(\tilde\Sbb),(n-1)\right).
\Eeq
Finally, we are able to express the first summation in term of $\Phi_{N,(n,m_1,m_2)}$:
\Beq
\sum_{i=1}^N\sum_{\Sbb}
\Delta^{\{\theta_\nu\}}_n
\Delta^{\{a_\nu\}}_{m_1}
\Delta^{\{b_\nu\}}_{m_2}
\theta_i(\Sbb)\phi_{N,\hatf_i\Sbb} 
=[N-(n-1)]\phi_{N,(n-1,m_1,m_2)}.
\Eeq
Proceeding along the same line for every summation symbol we have
\Beq
\frac{d}{dt}\Phi_{N,(n,m_1,m_2)} 
&=&
k
\left[1-\frac{(n-1)}{N}\right]\Phi_{N,(n-1,m_1,m_2)}
-
k
\left[1-\frac{n}{N}\right] \Phi_{N,(n,m_1,m_2)}\nonumber
 \\
&+&
\mu(n+1)
\Phi_{N,(n+1,m_1,m_2)} 
-
\mu n
 \Phi_{N,(n,m_1,m_2)}
  \\
&+& 
q_1 n 
  \Phi_{N,(n,m_1-1,m_2)} 
-
q_1 n   \Phi_{N,(n,m_1,m_2)}
  \nonumber\\
&+&
q_1 n 
 \Phi_{N,(n,m_1,m_2-1)} 
-
q_2 n \Phi_{N,(n,m_1,m_2)} 
\nonumber\\
&+&
\gamma_1[m_1+1]\Phi_{N,(n,m_1+1,m_2)} 
-
\gamma_1 m_1 \Phi_{N,(n,m_1,m_2)}
  \nonumber\\
&+& 
\gamma_2[m_2+1]
\Phi_{N,(n,m_1,m_2+1)} 
-
\gamma_2m_2\Phi_{N,(n,m_1,m_2)}.
 \nonumber
\Eeq
Taking the limit $N\rightarrow\infty$, we see that the latter equation converges towards the master equation \eqref{eq_Master_eq}. In other words 
$P_{n,m_1,m_2}=\lim_{N\rightarrow\infty}\Phi_{N;(n,m_1,m_2)}$, from which it naturally follow $G^{(0)}=\lim_{N\rightarrow\infty}\left(\Gi_{N}\right)^N$.

\subsection{Large time approximation}
The large time approximation, for one protein type only, is obtained by writting $r=\mu/\gamma$, $\delta=q(z-1)/\gamma$ and $\epsilon=e^{-\gamma t}$ so that
\Beq
(z-1)A=\delta\frac{k}{\gamma}\int_{\epsilon}^1\mathrm{d}u \int_u^1\mathrm{d}v \ \Omega(u,v),
\Eeq
with
\Beq
\Omega(u,v)=
\frac{u^{r-1}}{v^{r}}\exp\left[{\delta(z)(v-u)}\right].
\Eeq
Using the Taylor expansion leads to
\Beq
\label{eq104}
(z-1)A&=&\delta\frac{k}{\gamma}\sum_{n=0}^\infty\sum_{m=0}^\infty 
\frac{(-1)^m \delta^{n+m} }{m!n!}\int_{\epsilon}^1\mathrm{d}u \ u^{m+r-1} \int_u^1\mathrm{d}v \ v^{n-r}
,
\Eeq
for which there is no particular problem unless in the last integral we have $n-r=-1$ for some value of $n$. Avoiding this situation, by choosing $r\notin \mathbb{ N}$, leads to
\Beq
(z-1)A=I_F
&+&\delta\frac{k}{\gamma}
\sum_{n=0}^\infty\sum_{m=0}^\infty \frac{(-1)^m \delta^{n+m} }{m!n!}
\frac{1}{n-r+1}\nonumber
\\
&\times&
\left[\frac{1}{n+m+1}\epsilon^{n+m+1}
-\frac{1}{m+r}\epsilon^{m+r}\right],
\Eeq
with 
\Beq
I_F&=&\delta\frac{k}{\gamma}\sum_{\kappa=0}^\infty
\frac{\delta^{\kappa}}{\kappa+1}
\sum_{m=0}^\kappa \nonumber
\frac{(-1)^m }{m!(\kappa-m)!}\frac{1}{(m+r)}=\delta\frac{k}{\gamma}\sum_{\kappa=0}^\infty
\frac{\delta^{\kappa}}{\kappa+1}
\frac{(r-1)!}{(r+\kappa)!}\\
&=&\frac{k}{\mu}\int_0^\delta \mathrm{d}s \ _1F_1[1,r+1,s].
\Eeq
Keeping the lowest order in $\epsilon$ we get $(z-1)A\simeq I_F+h$ with 
\begin{equation}
h=\left\{
\begin{array}{cc}
\delta\frac{k}{\gamma}\frac{\epsilon}{1-r}& r>1\\
-\delta\frac{k}{\gamma}\frac{\epsilon^{r}}{r}\sum_{n=0}^\infty \frac{\delta^{n} }{n!}\frac{1}{n-r+1}& r<1.
\end{array}
\right.
\end{equation}
It follows that $G^{(0)}(z,t)\simeq G^*(z)H(z,t)$, with $H=e^{h}$ as presented in equation \eqref{eq_HH}. Going back to equation \eqref{eq104}, we can work with $r=1$ ($\mu=\gamma$). In this case, when keeping terms of order $\epsilon$ and $\epsilon\ln(\epsilon)$ we get :
\begin{equation}
h= -\delta\frac{k}{\mu}e^{-\mu t}\left(\mu t+\sum_{n=1}^\infty\frac{\delta^n}{n\times n!}\right).
\end{equation}
Under the following approximation we get $\moyenne{m}(t)/\moyenne{m}^*\simeq1-\mu te^{-\mu t}$ in agreement with the long time limit of equation \eqref{eq5}.

\end{document}